\documentclass[twocolumn,showpacs,prb,amsfonts,amsmath,amssymb,floatfix,groupedaddress]{revtex4-1}
\usepackage{color}
\usepackage{hhline}
\usepackage{mathrsfs}
\usepackage{graphicx}
\usepackage{dcolumn}
\usepackage{bm}% bold math
\usepackage{multirow}
\usepackage{booktabs}
\usepackage{afterpage}
\usepackage{amsmath}
\usepackage{ulem}

\begin{document}

%\title{Why is the DOS peak robust in cubic H$_{3}$S?}
%\title{Band-topography enforced DOS peaking in three-dimensional crystals: Theory and case study of cubic H$_{3}$S}
\title{Archetypical ``push the band critical point" mechanism for peaking of the density of states in three-dimensional crystals: Theory and case study of cubic H$_{3}$S}
\author{Ryosuke Akashi$^{1}$}
\thanks{akashi@cms.phys.s.u-tokyo.ac.jp}
\affiliation{$^1$Department of Physics, The University of Tokyo, Hongo, Bunkyo-ku, Tokyo 113-0033, Japan}

\date{\today}
\begin{abstract}
The point of zero gradient of the electronic band structure--critical point--generally induces the singularity in the density of states (DOS), but no isolated critical point yields strict divergence of the DOS in three dimensions, differently from the lower dimensional cases. In view of the band structure as a smooth hypersurface on the reciprocal space, we discuss the minimal deformation of the band structure that yields non-divergent but large sharp DOS peaks in three dimensions. By ``pushing down" the energy level at the second order saddle point (maximum), a continuous closed loop of saddle points (sphere of maxima) encircling the original position of the saddle point (maximum) emerges, with which the DOS peak is formed. Being high dimensional features, the saddle loop and extremum shell thus formed are difficult to locate with standard band structure analysis on linear ${\bf k}$-point paths. The Lifshitz transition occurring over a linear or planar manifold is discussed as an indicator of such features. We also find that the celebrated DOS peak in the recently discovered superconducting hydride H$_{3}$S originates from the saddle loop. On this basis, we successfully extract the minimal model that explains how the DOS peak is formed. Our theory characterizes a large class of DOS peaks sometimes found in the three dimensional electronic structures, building a basis for profound understanding of their origins.
\end{abstract}

\maketitle

\section{Introduction}
The diversity of the electronic property of condensed matter wholly originates from the ionic configuration. One of its most striking consequences is anomalous concentration of the electronic one-particle states characterized by the peaks in the spectrum of the density of states (DOS) $D(E)$. The concentration of the DOS near the Fermi level indicates that many electrons contribute to the low-energy phenomena, as well as implies that effects of the electron-electron and electron-ion interactions become significant. 

How to make the DOS concentrate in tiny energy ranges has therefore been of continuous interest in the field of the band theory. An extreme example is the flat band,~\cite{Sutherland1986, Lieb1989,Mielke1-1991,Mielke2-1991,Tasaki1992, Mielke-Tasaki1993, Tasaki-review, Bergholtz-review} where the electronic one-particle energy eigenvalues $\varepsilon({\bf k})$ with ${\bf k}$ being the Bloch wave number is constant in the entire Brillouin zone. Theoretically, the flat bands emerge from tight binding models with specially designed features such as decoration of the unit cell and carefully tuned model hopping parameters between the orbitals. A recent interesting realization of almost flat band is seen in the twisted bilayer graphene.~\cite{TBG1,TBG2,MacDonald2012} Band structures with flat dispersions within partial regions of the ${\bf k}$ space and the resulting divergence (or peak) in $D(E)$ are often reported; we list a few which have been understood as a consequence of the special configurations of the hopping parameters.~\cite{Ochi-RCo5-2015, Jelitto1969, Akashi-interfere-PRB2017}
%In most cases, the exact flat band is ensured only when the model of the system strictly satisfies strong constraints that cannot be determined only by the spatial symmetry. Suppose the electronic system is modeled by the tight-binding Hamiltonian $\mathcal{H}=\sum_{ij}t_{ij}c^{\dagger}_{i}c_{j}$ [$c^{\dagger}_{i} (c_{i})$: creation (annihilation) operator of state $i$]. To the author's knowledge, the known mechanisms of the flat band formation requires either special configuration of the states $i$ or relations between the components of $t_{ij}$. 

A theory from a contrasted viewpoint, where we only assume differentiability of $\varepsilon({\bf k})$ in the ${\bf k}$ space, has been established by van Hove.~\cite{vHS} He has pointed out that, the critical points defined by $|\nabla \varepsilon({\bf k})|=0$ always yield singular points in the DOS (van Hove singularity; vHS) and the minimum number of critical points is nonzero due to the topology of the ${\bf k}$ space. Remarkably, in the one and two dimensional cases, divergent singularities in $D(E)$ related to the isolated critical points always emerge. The saddle point in (quasi-) two dimensional systems has therefore been of recurrent interest in different contexts; in cuprates,~\cite{PhysRevLett.73.3302, Markiewicz-review} topological surface states,~\cite{PhysRevB.97.075125,Ghosh-Pt2HgSe3-arxiv} and the monolayer and multilayer graphene,~\cite{PhysRevB.77.113410, PhysRevLett.104.136803, Chubukov-graphene-SC, PhysRevB.95.035137, Yuan-TBLG-arxiv} etc. However, such divergence is not generally assured to be present in the three-dimensional case. The isolated critical point with the quadratic {\bf k} dependence induces the divergence of the derivative of the DOS, but not of the DOS value. The peaks in the DOS are nevertheless observable in various three dimensional systems with the model and first principle electronic structure calculations, even if dimensional reduction is not apparent from the crystal structure of the systems. This fact motivates us to seek for commonplace deformations to $\varepsilon({\bf k})$ that can yield divergent DOS singularities in three dimensions.

In the present article, we characterize a general but unrecognized mechanism of forming divergent singularities in $D(E)$ in three dimensions. Energetically degenerate continuous extension of the critical points generally enhances the degree of singularity by reducing the effective dimension.~\cite{Markiewicz-review} We point out that, by ``pushing" the isolated critical point on the hypersurface $\varepsilon({\bf k})$, nearly degenerate loop or shell structure of new critical points are formed around it, giving rise to the DOS peaks. We discuss the archetype of this mechanism and how it appears in reality with a simple model. 
%When such DOS peaks are found to emerge, some of them have been found to be consequences of the band flattening in some directions induced by special geometrical relations of $t_{ij}$, but mostly the mechanisms of their formation are left unexplored. 
Also, we show that our theory successfully describes how the DOS peak forms in the recently discovered pressure induced high-$T_{\rm c}$ superconductor H$_{3}$S,~\cite{Eremets} which boosts its $T_{\rm c}$ up to 200~K by incorporating a large number of electrons into the pair condensation.

\section{Theory}

\begin{figure}[h]
    \begin{center}
        \includegraphics[scale=0.30]{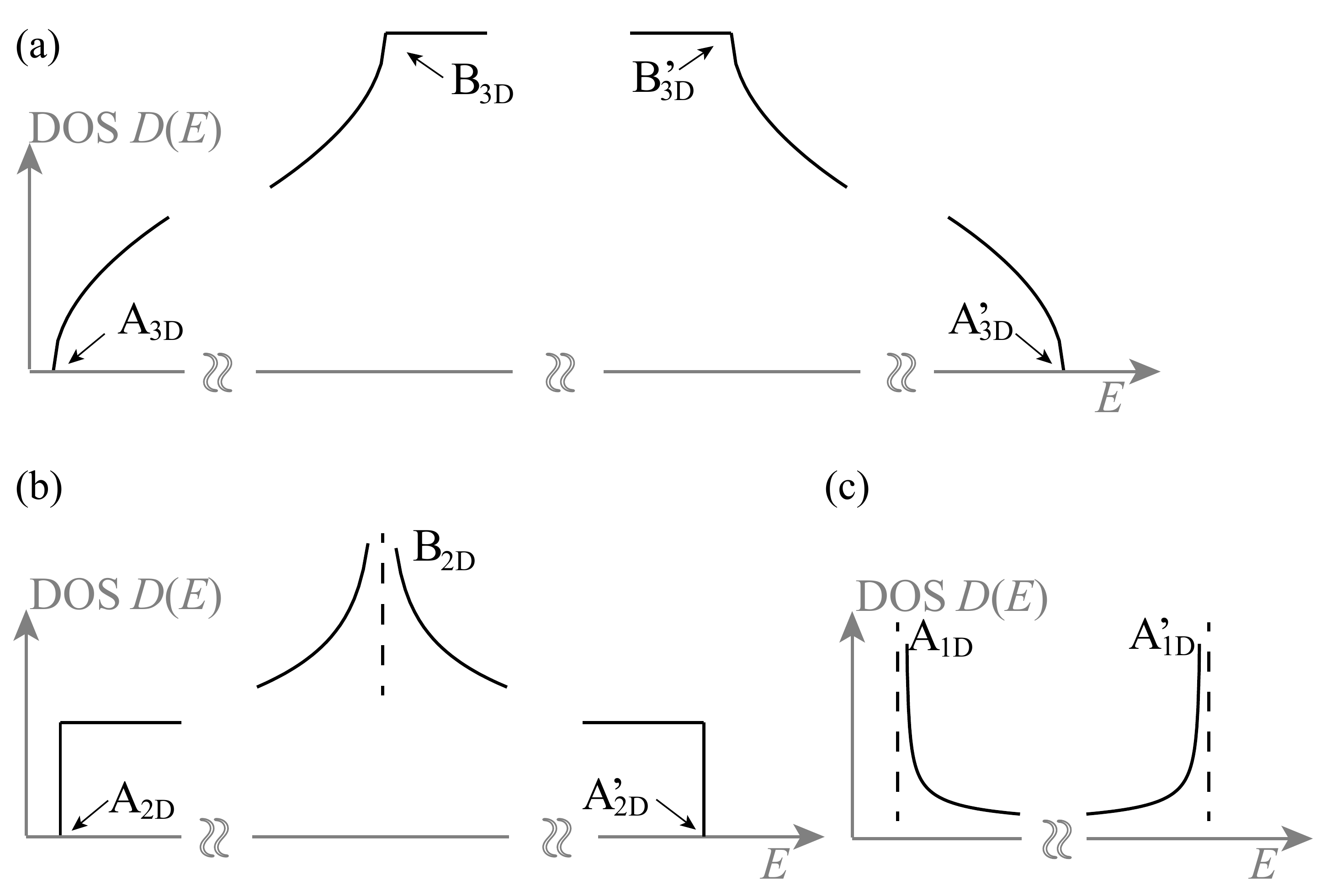}
        \caption{Asymptotic forms of the DOS near the vHSs for (a) three, (b) two, and (c) one dimensions. $A (A')$ and $B (B')$ represent the vHSs corresponding to the extrema and saddle points, respectively. See the main text for more specific definitions of the vHSs.}
        \label{fig:vHSclass}
    \end{center}
\end{figure}

\subsection{Critical points and the van Hove singularities}
Let us first start from the review on the relation between the critical points and vHSs in the electronic DOS. The critical point is defined as the points ${\bf k}$ which satisfy $|\nabla \varepsilon({\bf k})|=0$. The band dispersion near the critical point is therefore generally approximated to the following bilinear form
\begin{eqnarray}
\varepsilon_{\rm 3D}({\bf k})
&\simeq &
\frac{1}{2}
\left(
\begin{array}{ccc}
k_{x} &\ k_{y} &\ k_{z}
\end{array}
\right)
\left(
\begin{array}{ccc}
a_{xx} &\  a_{xy} &\  a_{xz} \\
a_{xy} &\  a_{yy} &\  a_{yz} \\
a_{xz} &\  a_{yz} &\  a_{zz}
\end{array}
\right)
\left(
\begin{array}{c}
k_{x} \\
k_{y} \\
k_{z}
\end{array}
\right)
\nonumber \\
&=&
\frac{k^{2}_{1}}{2m_{1}}
+
\frac{k^{2}_{2}}{2m_{2}}
+
\frac{k^{2}_{3}}{2m_{3}}
,
\label{eq:bilinear3D}
\end{eqnarray}
\begin{eqnarray}
\varepsilon_{\rm 2D}({\bf k})
&\simeq &
\frac{1}{2}
\left(
\begin{array}{cc}
k_{x} &\ k_{y}
\end{array}
\right)
\left(
\begin{array}{cc}
a_{xx} &\  a_{xy}  \\
a_{xy} &\  a_{yy}  
\end{array}
\right)
\left(
\begin{array}{c}
k_{x} \\
k_{y}
\end{array}
\right)
\nonumber \\
&=&
\frac{k^{2}_{1}}{2m_{1}}
+
\frac{k^{2}_{2}}{2m_{2}}
,
\\
\varepsilon_{\rm 1D}({\bf k})
&\simeq &
\frac{k^{2}_{1}}{2m_{1}}
,
\end{eqnarray}

where $(k_{1}, k_{2}, k_{3})^{T} \equiv M_{\rm 3D} (k_{x}, k_{y}, k_{z})^{T}$ [$(k_{1}, k_{2})^{T} \equiv M_{\rm 2D} (k_{x}, k_{y})^{T}$] with $M_{\rm 3D}$ ($M_{\rm 2D}$) being an orthogonal matrix. The structure in the DOS corresponding to the vHS is classified by the relations between the values of the effective mass $m_{i}$. Let us define the types of the critical points for the three dimensional case as 
\begin{eqnarray}
&&A_{\rm 3D}: m_{1} > 0,\  m_{2} > 0,\  m_{3} > 0,
\\
&&B_{\rm 3D}: m_{1} > 0,\  m_{2} > 0,\  m_{3} < 0,
\\
&&B'_{\rm 3D}: m_{1} > 0,\  m_{2} < 0,\  m_{3} < 0,
\\
&&A'_{\rm 3D}: m_{1} < 0,\  m_{2} < 0,\  m_{3} < 0,
\end{eqnarray}
for the two dimensional case as
\begin{eqnarray}
&&A_{\rm 2D}: m_{1} > 0,\  m_{2} > 0,
\\
&&B_{\rm 2D}: m_{1} > 0,\  m_{2} < 0,
\\
&&A'_{\rm 2D}: m_{1} < 0,\  m_{2} < 0,
\end{eqnarray}
and for the one dimensional case as
\begin{eqnarray}
&&A_{\rm 1D}: m_{1} > 0,
\\
&&A'_{\rm 1D}: m_{1} < 0,
\end{eqnarray}

Figure 1 represents the typical behavior of the DOS near the vHS related to the critical points in the respective dimensions.~\cite{Grosso-Parravicini} In the three dimensional case, the critical points are classified into two: (i) all the effective mass have the same sign (extremum; $A_{\rm 3D}, A'_{\rm 3D}$), or (ii) either of the three has the different sign (saddle point; $B_{\rm 3D}, B'_{\rm 3D}$). Near the vHSs related to those critical points (commonly represented by $E_{\rm vHS}$ below), the DOS behaves
\begin{eqnarray}
D(E)
&\propto&
\sqrt{E-E_{\rm vHS}} \  (A_{\rm 3D})
\\
D(E) &\propto& 
\left\{
\begin{array}{c}
-\sqrt{E_{\rm vHS}-E}+{\rm const.}  \ (E < E_{\rm vHS})\\
{\rm const.} \hspace{65pt} (E > E_{\rm vHS})
\end{array}
\right.
 (B_{\rm 3D})
 .
\end{eqnarray}
The classification for two dimensional case is as follows: (i) all the effective mass have the same sign (extremum; $A_{\rm 2D}, A'_{\rm 2D}$), or (ii) the signs of the values of the effective mass are different (saddle point; $B_{\rm 2D}$).  
 \begin{eqnarray}
D(E)
&\propto&
\theta(E-E_{\rm vHS}) \  (A_{\rm 2D})
\\
D(E) &\propto&
{\rm log}\frac{1}{|E_{\rm vHS}-E|} 
\  (B_{\rm 2D})
 ,
\end{eqnarray}
where $\theta$ denotes the theta function $\theta(x)=0 \ (x<0); 1 (x>0)$. The dependence related to $A'_{\rm 2D}$ is inverse of $A_{\rm 2D}$ as well. In the one dimensional case, the one effective mass is either positive ($A_{\rm 1D}$) or negative ($A'_{\rm 1D}$)
\begin{eqnarray}
D(E)
&\propto&
\frac{1}{\sqrt{E-E_{\rm vHS}}} \  (A_{\rm 1D})
\end{eqnarray}
The dependences related to $A'_{\rm 3D}$, $B'_{\rm 3D}$, $A'_{\rm 2D}$ and $A'_{\rm 1D}$ are inverse of those without primes, respectively. In later discussions we often focus only on the critical points with (without) prime, as parallel arguments are obviously applicable to those without (with) prime by the energy inversion.

\begin{figure}[h]
    \begin{center}
        \includegraphics[scale=0.35]{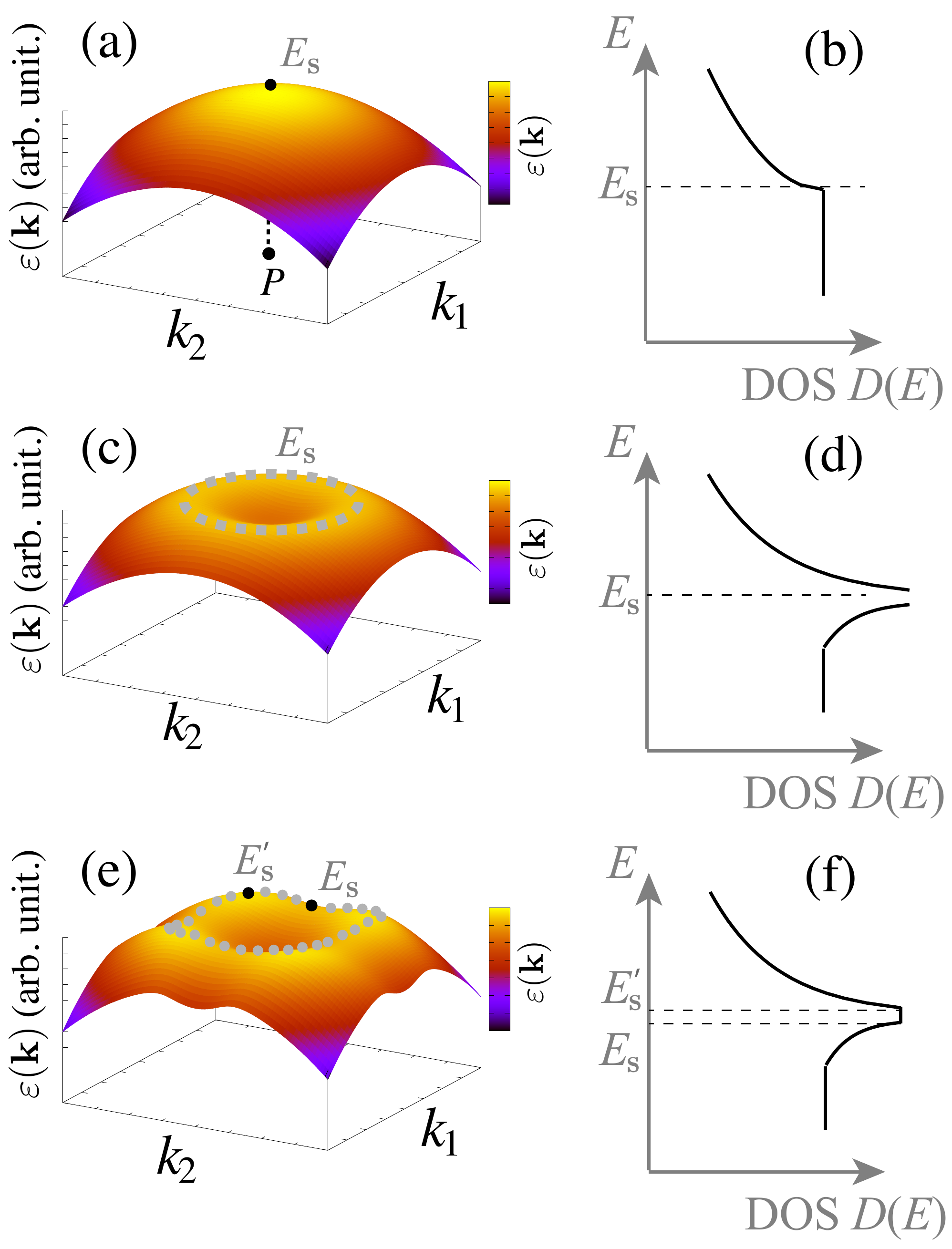}
        \caption{Schematic picture of the concept of perturbing the saddle point $P$ in three dimensions. (a) Two dimensional band structure near the original saddle point and (b) the corresponding form of the DOS near its energy level $E_{\rm s}$. In the third dimension (not shown) the band is assumed parabolic. (c) The band structure with the circular symmetric perturbation turning $P$ from the maximum to minimum in the two dimensions. The dashed line indicates the saddle loop. (d) Form of the DOS, where the divergence occurs at the energy level of the saddle loop. (e) The band structure with anisotropic deformation to (c), where we indicate the ridge as a remnant of the saddle loop as dotted line. The emergent saddle points are denoted by $E_{\rm s}$ and $E'_{\rm s}$. (f) The corresponding schematic form of the DOS, where shoulders are formed at $E_{\rm s}$ and $E'_{\rm s}$, respectively.}
        \label{fig:caldera}
    \end{center}
\end{figure}

\begin{figure}[h]
    \begin{center}
        \includegraphics[scale=0.35]{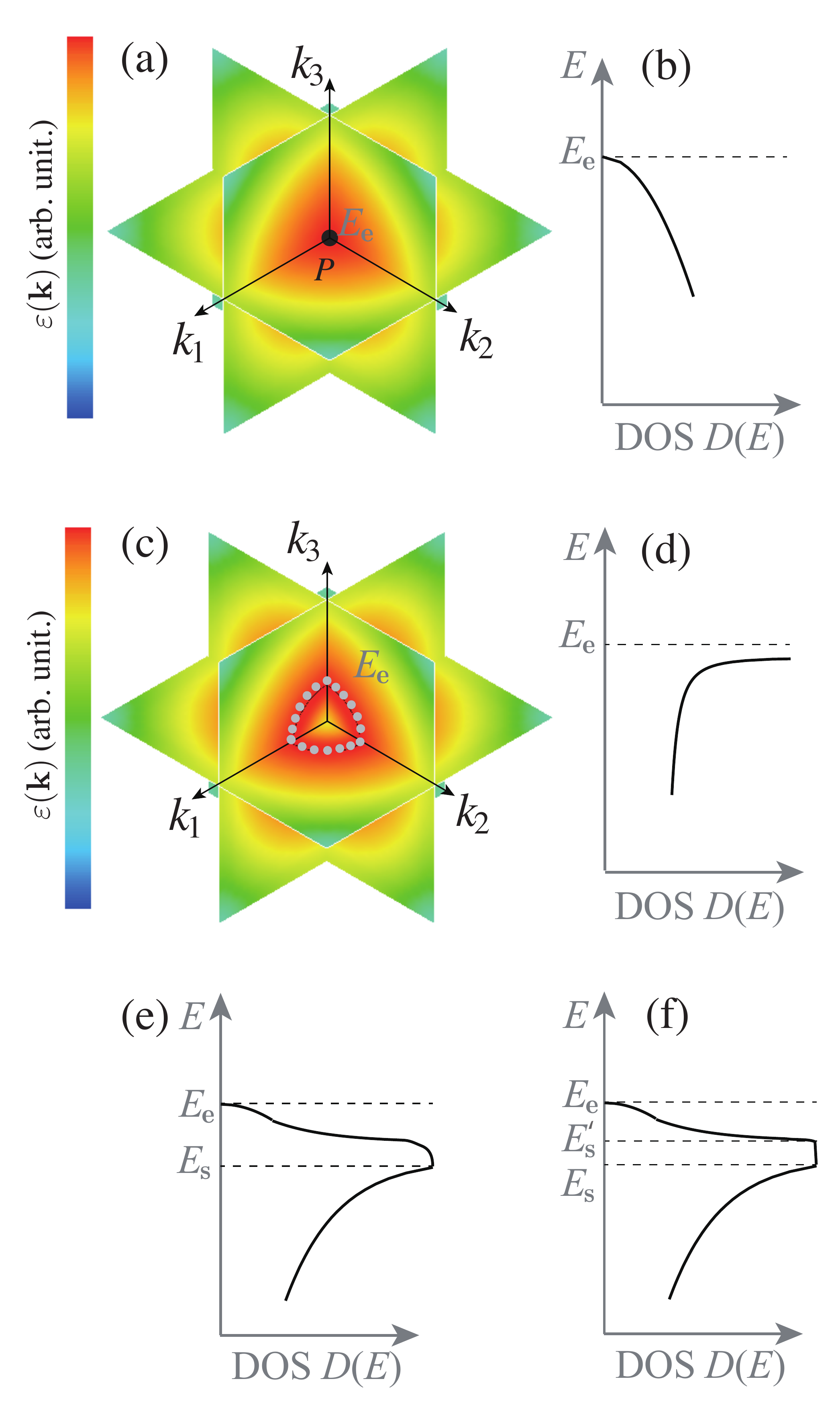}
        \caption{Schematic picture of the concept of perturbing the extremum $P$ in three dimensions. (a) Cross sections of the three dimensional band structure near the original maximum. Here, the value of $\varepsilon({\bf k})$ is represented by the color scheme. (b) The corresponding form of the DOS near its energy level $E_{\rm e}$. (c) The band structure with the spherically symmetric perturbation turning $P$ from the maximum to minimum. The dotted line indicates the extremum shell. (d) The form of the DOS, where the divergence occurs at the energy level of the extremum shell $E_{\rm e}$. (e) The schematic form of the DOS for the case with anisotropic deformation to (c), where termination point corresponding to the maximum $E_{\rm e}$ and the shoulder corresponding to the saddle point $E_{\rm s}$ are formed, respectively (The corresponding three dimensional band structure is not shown). (f) The variant of (e) with additional saddle point $E'_{\rm s}$ on the shell, where the corresponding shoulder is formed.}
        \label{fig:sphere}
    \end{center}
\end{figure}

The above discussion clarifies that, for possible divergence of DOS in three dimensions, it is necessary that the bilinear expansion should be broken down at certain band critical points. In other words, the rank of the coefficient matrix in Eq.~(\ref{eq:bilinear3D}) must be lower than 3 (Ref.~\onlinecite{Yuan-TBLG-arxiv}). This means that the expansion series in certain coordinates begin with the higher order terms, or, as a limiting case, the critical point extends over a continuous one or higher dimensional region in the {\bf k} space. For simplicity we ignore the higher order terms and concentrate on the latter possibility in this paper; the band dispersion is either parabolic or flat. From this view, by taking the limits where any of the effective masses becomes infinity, the form of the DOS around the vHS converges to any low-dimensional counterpart. For example, for the $B_{\rm 3D}$ case, in the limit $m_{2}\rightarrow \infty$ the DOS form converges to that of $B_{\rm 2D}$, whereas the limit $m_{3}\rightarrow \infty$ corresponds to $A_{\rm 2D}$. These correspondences reflects that the effective dimension of the band structure is reduced due to the extension of the critical point. The higher order components of $\varepsilon({\bf k})$ can make subtle changes to the degree of singularity, though we ignore them. What we do in this work is consider a simple modification of general three dimensional band structures which yields the extended critical points for possible utilization of the diverging nature of $D(E)$ in lower dimensions.

\subsection{Saddle loop and extremum shell}
In this subsection, we characterize a general mechanism for emergence of the peak in the DOS in three dimensional system. Let us consider a saddle point $P$ of class $B'_{\rm 3D}$ for example (Fig.~\ref{fig:caldera}(a), left), where in the $k_{1}$ and $k_{2}$ directions the band dispersion is concave ($m_1 = m_2 <0$). We assume that the dispersion is kept parabolic in the $k_{3}$ direction with positive definite effective mass around $P$ in the equal-$k_{3}$ plane ($m_{3}>0$) in the later discussion in this paragraph. In this case, the resulting DOS exhibit the shoulder like vHS [Fig.~\ref{fig:caldera}(b)] corresponding to the critical point $B'_{\rm 3D}$ in Fig.~\ref{fig:vHSclass}. Suppose any perturbation is exerted on the system, which significantly reduces the energy eigenvalues in the close vicinity of $P$ (push the $P$ point). With sufficiently large amount of the perturbation, $P$ turns into the energy minimum and caldera-like structure emerges. If the change of the energy eigenvalues is ideally circular symmetric with respect to the $k_{1}$--$k_{2}$ plane, a closed loop of the saddle points, or saddle loop, appears as depicted in Fig.~\ref{fig:caldera}(c). At all the points on this line, the band dispersion is concave in the direction perpendicular to the line, flat along the line, and convex in the $k_{3}$ direction as assumed above. As a result, a divergent singularity of the DOS corresponding to the case $B_{\rm 2D}$ is formed [Fig.~\ref{fig:caldera}(d)]. Interestingly, the original saddle point $P$ is then modified into the local minimum $A_{\rm 3D}$ and it does not appear in the DOS spectrum since the contribution around the point, being proportional to $\sqrt{E-E_{P}}$, is overwhelmed by the contribution from the vicinity of the saddle loop. 

One can also conceive a similar situation for the maximum point $P$ corresponding to type $A'_{\rm 3D}$. For simplicity let us assume the band isotropy around this point as well ($m_{1}=m_{2}=m_{3}<0$). The DOS form near the singularity trivially corresponds to $\propto \sqrt{E_{P}-E}$ [Fig.~\ref{fig:sphere} (a)(b)]. By introducing a perturbation with spherical symmetry that appreciably reduces the energy eigenvalues around $P$, the maxima of the resulting spectrum form a closed shell, or extremum shell, around $P$[Fig.~\ref{fig:sphere}(c)]. At any points on this shell, the band dispersion is zero and parabolic in the two tangential directions and one perpendicular direction, respectively. The divergent singularity of the $A'_{1D}$ type then appears at the energy value on the shell [Fig.~\ref{fig:sphere}(d)].

\begin{figure}[h]
    \begin{center}
        \includegraphics[scale=0.21]{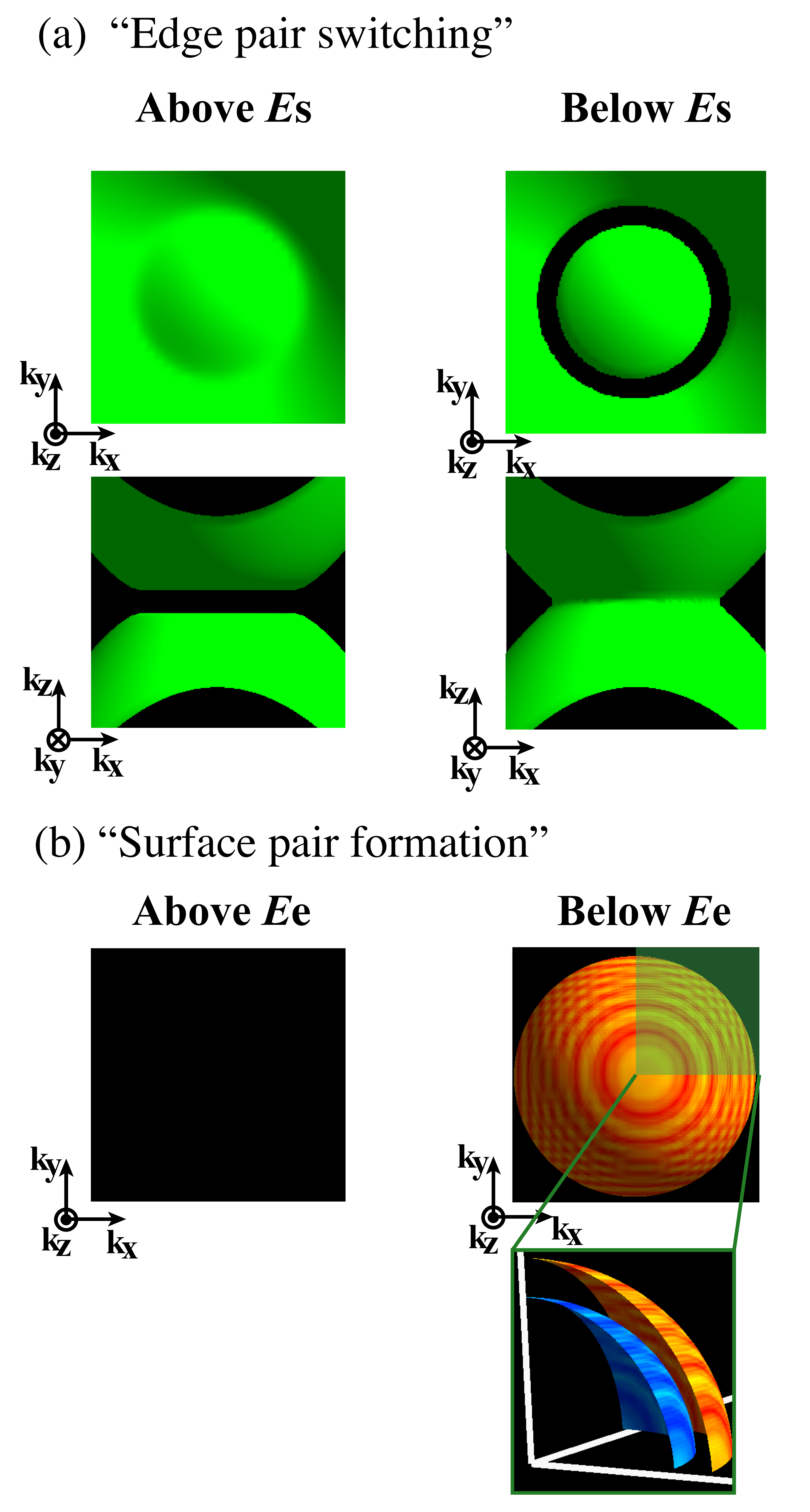}
        \caption{Higher order Lifshitz transitions. (a) The edge pair switching transition in the equal energy surface, where the surfaces at the energy levels slightly above and below $E_{\rm s}$ for the band structure of Fig.~\ref{fig:caldera}(c). (b) The surface pair formation transition, where the surfaces at the energy levels slightly above and below $E_{\rm e}$ for the band structure of Fig.~\ref{fig:sphere}(c). The close up view of the cross sections is also displayed, where the white lines indicate the cutting planes.}
        \label{fig:LTs}
    \end{center}
\end{figure}

\subsubsection{Higher order Lifshitz transitions}
We here show the character of the saddle loop and extremum shell from the perspective of the equal energy surface. The equal energy surface is defined by the manifold in the ${\bf k}$ space where $\varepsilon({\bf k})$ is equal to a certain energy value. The surface at the Fermi energy--Fermi surface--is of particular interest as it displays the electronic states dominating low-energy phenomena. The change of the topology of the equal energy surface by varying the energy value is called Lifshitz transition.~\cite{Lifshitz} In three dimensions, generally two types of the transition occur: pocket appearance and the neck disruption. These transitions are observed when the energy value varies through the levels of critical points of types $A_{\rm 3D}$ and $B_{\rm 3D}$, respectively. When we change the energy value through the vHS due to the saddle loop or extremum shell, topologically distinct transitions occur in the equal energy surface: so to speak, (i) the edge pair switching transition~[Fig.~\ref{fig:LTs}(a)]--two adjacent surfaces are attached along the saddle loop and a pocket is detached with the switching of the edge connections--and (ii) surface pair formation transition~[Fig.~\ref{fig:LTs}(b)]--two parallel surfaces emerge along the extremum shell from the void. Those transitions are actually variants of the known neck disruption transition, in that they are understood as the simultaneous occurrence of the neck disruption (or formation) on a one and two dimensional manifolds, respectively. For this reason we later refer to those transitions as higher-order Lifshitz transitions. We propose to give them specific names, since those transitions are directly related to the source of the divergence of DOS; namely, the linear (planar) region in the ${\bf k}$ space where ``edge pair switching (surface pair formation)" occurs is responsible for the divergent contribution. Note that the neck disruption at a point indicate the divergence of the derivative of DOS, but not of the DOS itself.

\subsubsection{Effect of anisotropy}
In realistic three dimensional systems, we cannot generally expect the ideal saddle loop and extremum shell since they require that the gradient of the band is exactly zero along them. Still, we can state that the DOS peaks resulting from those, though not divergent, persist against subtle deformations of the band structure, as long as the the ${\bf k}$ dependence of $\varepsilon({\bf k})$ is not drastic. In the above discussions of the saddle loop, we have assumed circular symmetry for the perturbation around the original saddle point $P$. When some anisotropy is introduced, the saddle loop is deformed into a (not ideally circular) loop of ridge line encircling $P$ [Fig.~\ref{fig:caldera}(e)], along which the band dispersion is not exactly zero. In the middle of this line, there must emerge one or more minima and maxima, which corresponds to the critical points of type $B_{\rm 3D}$ and $B'_{\rm 3D}$, respectively. The divergent DOS singularity then evolves into two (or more) neighboring shoulders, whose energy distance corresponds to the band dispersion along the ridge loop. The whole structure appears as a ``peak" in the calculated DOS. Note that the presence of the two types of critical points is due to the periodicity of $\varepsilon({\bf k})$ along the loop, and they are not assured for $\varepsilon({\bf k})$ on an open linear region.

A parallel discussion applies to the case of the extremum shell. With the deformation of the band structure from the above ideal case, the shell is somehow deformed and at least two critical points respectively of $A'_{\rm 3D}$ and $B_{\rm 3D}$ are assured to appear anywhere on the shell, and can $B'_{\rm 3D}$ appear occasionally. The divergent DOS singularity then evolves into adjacent termination point and shoulder(s), between which the DOS rapidly varies [Fig.~\ref{fig:sphere} (e)(f)]. Note that the apparent width of the DOS peak does not exactly corresponds to the energy distance between the $A'_{\rm 3D}$ and $B_{\rm 3D}$ points [$E_{\rm e}-E_{\rm s}$ in panel (f)]. 

The anisotropy also changes the appearance of the higher order Lifshitz transitions depicted above. When the energy level is tuned through the peak positions, the transition initiates from the isolated vHSs anywhere on the loop or shell. The starting and end points of the transitions would appear as the well-known neck disruption or formation. By comparing the equal energy surfaces below the lower vHS and above the higher vHS, we see the edge pair switching or surface pair formation process. A statement therefore remains valid that, even with the anisotropy, the higher order Lifshitz transitions occurring between a certain (nonzero) energy range indicate the peaked concentration of the DOS within that range, as well as displays the ${\bf k}$ point region responsible for the peak.

In the literature, one sometimes finds discussions relating the peak of the DOS and any vHS appearing in the band structure along specific linear paths. For the above two cases, the DOS peaks are better understood as remnant of the ideal saddle loop or extremum shell. The total DOS peak is formed by the states near the loop or shell, not only by those in the vicinity of the isolated vHSs. The vHSs would correspond to any detailed feature of the peak such as shoulders, but what dominates the peak width is the degree of the band dispersion over the loop or shell, where specific positions of the vHS could change depending on subtle perturbations. We believe that many of the DOS peaks found so far in the model and first-principles band structure calculations of the three dimensional systems should be of the present classes. Later, we demonstrate a simple model system that exhibit the DOS peaks due to the saddle loop and extremum shell and analyze the first-principle band structure of a realistic system, superconducting H$_3$S under extreme compression,~\cite{Eremets} from the viewpoint of the present theory.

\section{Tight binding model}
In the previous section, we have discussed the possible DOS peaks in three dimensions due to formation of the saddle loops and extremum shell. This mechanism is apparently so general that one would expect it is ubiquitous, though the requirements for its occurrence seems difficult to realize in terms of the Hamiltonian. The essential factor is that the perturbation acts strongly only at the close vicinity of the original saddle point or extremum in the ${\bf k}$ space. Another factor in the case of the saddle loop is that the band dispersion in the other direction must remain parabolic with definite sign of the effective mass. Here we exemplify a simple model system where those requirements are satisfied.

\begin{figure}[t]
    \begin{center}
        \includegraphics[scale=0.30]{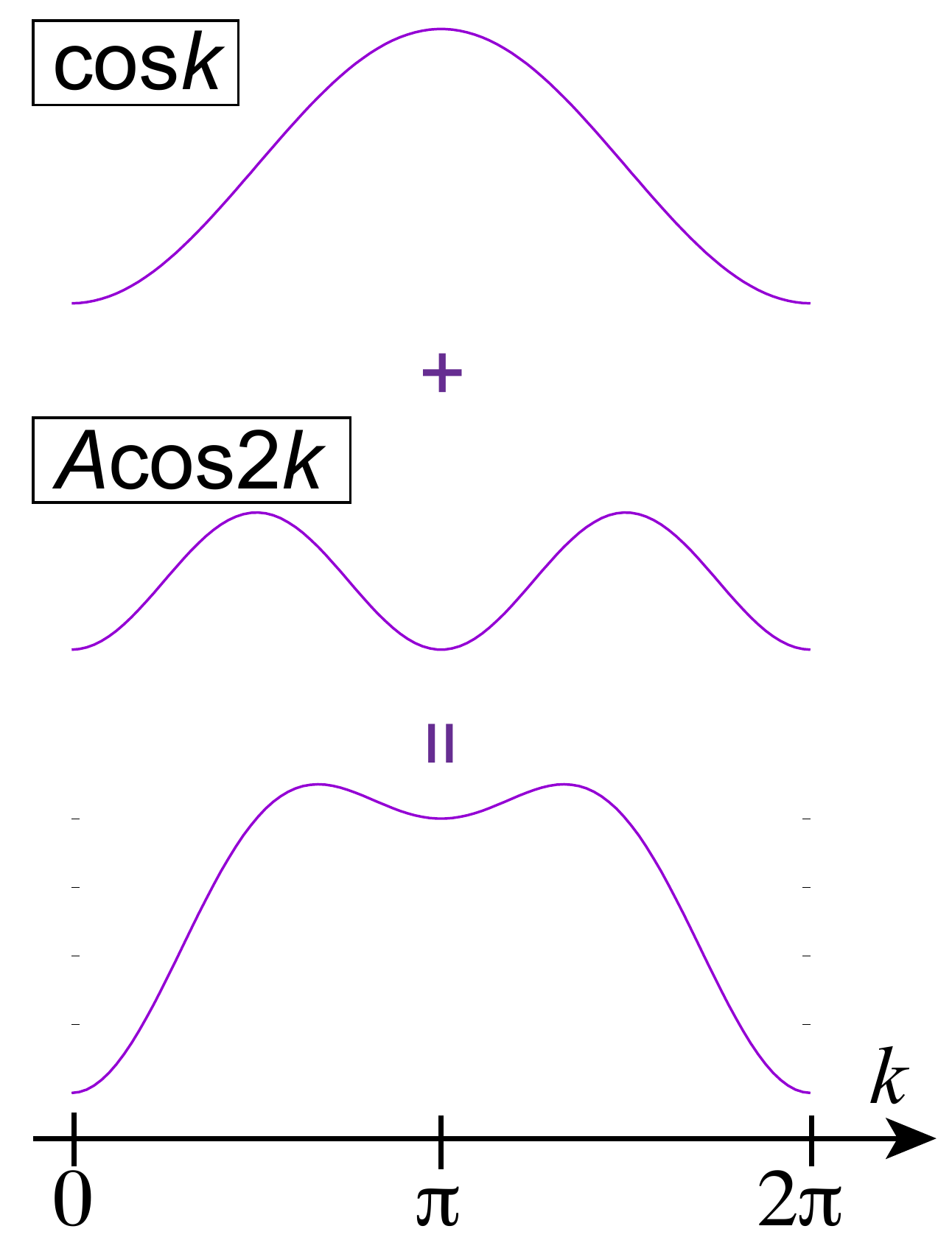}
        \caption{Modification of the maximum (at $k=\pi$) to minimum by a rapidly varying Fourier component. $A$ denotes a parameter.}
        \label{fig:cos-cos2}
    \end{center}
\end{figure}

\subsection{Promising perturbations}
Before going to the specific model, we argue possible perturbations in terms of the tight-binding model that can yield the saddle loops and extremum shells. The tight-binding model on a periodic lattice is generally written as
\begin{eqnarray}
\mathcal{H}
=\sum_{{\bf R}, {\bf R}'}\sum_{ij} t_{ij}({\bf R}-{\bf R}')c_{i}^{\dagger}({\bf R})c_{j}({\bf R}')
.
\end{eqnarray}
Here, ${\bf R}$ and ${\bf R}'$ denote the lattice vectors and $i$ and $j$ are the index of orbitals.  $c^{\dagger}_{i}({\bf R})$ ($c_{i}({\bf R})$) is the creation (annihilation) operator of electrons of state $i$ at site ${\bf R}$. $t_{ij}({\bf R})$ is the hopping between states $i$ and $j$ across the unit cells connected by vector ${\bf R}$. The hopping with ${\bf R}$ yields the ${\bf k}$ dependence of the energy eigenvalues $\varepsilon_{n}({\bf k})$ ($n:$ band index) with the form $\sim {\rm cos}({\bf k}\cdot {\bf R})$; the near-site hopping (=small ${\bf R}$) thus gives slowly varying component in the ${\bf k}$ space. If the band structure near the saddle or maximum point is dominated by such slowly varying components, it is modified to the minimum by adding more rapidly oscillating contributions which originates from the hopping with larger ${\bf R}$. In Fig.~\ref{fig:cos-cos2}, we depict an example: In one direction, the maximum of the dispersion $\sim {\rm cos}k$ at $k=\pi$ is modified to the minimum by adding the component with the ${\rm cos}2k$ form. Note that in this case the relative amount of the ${\rm cos}2k$ component represented by parameter $A$ in Fig.~\ref{fig:cos-cos2} must be larger than some threshold value in order to change the maximum at $k=\pi$ to local minimum. 

In multiband systems, there can also be a possibility that the composition of the states labeled by $n{\bf k}$ varies rapidly with respect to ${\bf k}$, as observed in the topological insulator.~\cite{RevModPhys.83.1057} In such a situation, one could render the target saddle or maximum to minimum point by exerting any orbital-selective perturbation.

The farther neighbor hopping is especially promising as the candidate perturbation in that it is generally expected to become appreciable in systems under strong compression. Below, we first demonstrate the formation of the DOS peak in a tight-binding model due to the farther neighbor hopping. We later discuss the origin of the DOS peak at the Fermi level in the compressed H$_{3}$S, where we find that the perturbation responsible for the peak formation is the farther neighbor orbital selective hopping.

\begin{figure}[h]
    \begin{center}
        \includegraphics[scale=0.50]{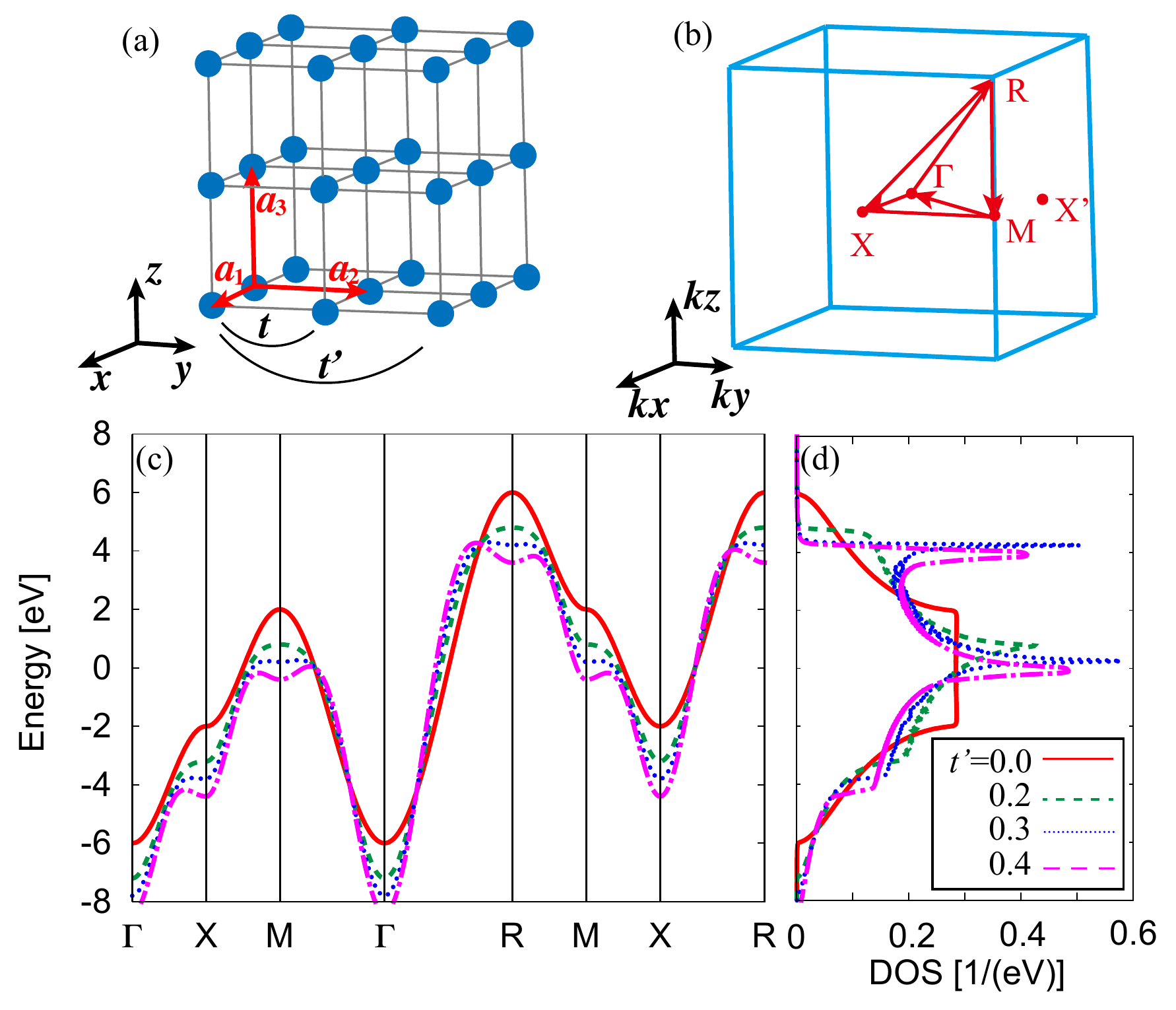}
        \caption{(1) Simple cubic tight binding model with farther neighbor hopping. (b) The simple cubic Brillouin zone, where the ${\bf k}$-point path for the band structure calculation is depicted. The point $X'$ is a symmetrically equivalent to $X$. (c) The calculated band structure and (b) the DOS of the model of the spectrum Eq.(\ref{eq:ek-sc-tp}). Here and hereafter $t$ is fixed to 1~eV.}
        \label{fig:sc-band}
    \end{center}
\end{figure}

\begin{figure*}[t!]
    \begin{center}
        \includegraphics[scale=0.32]{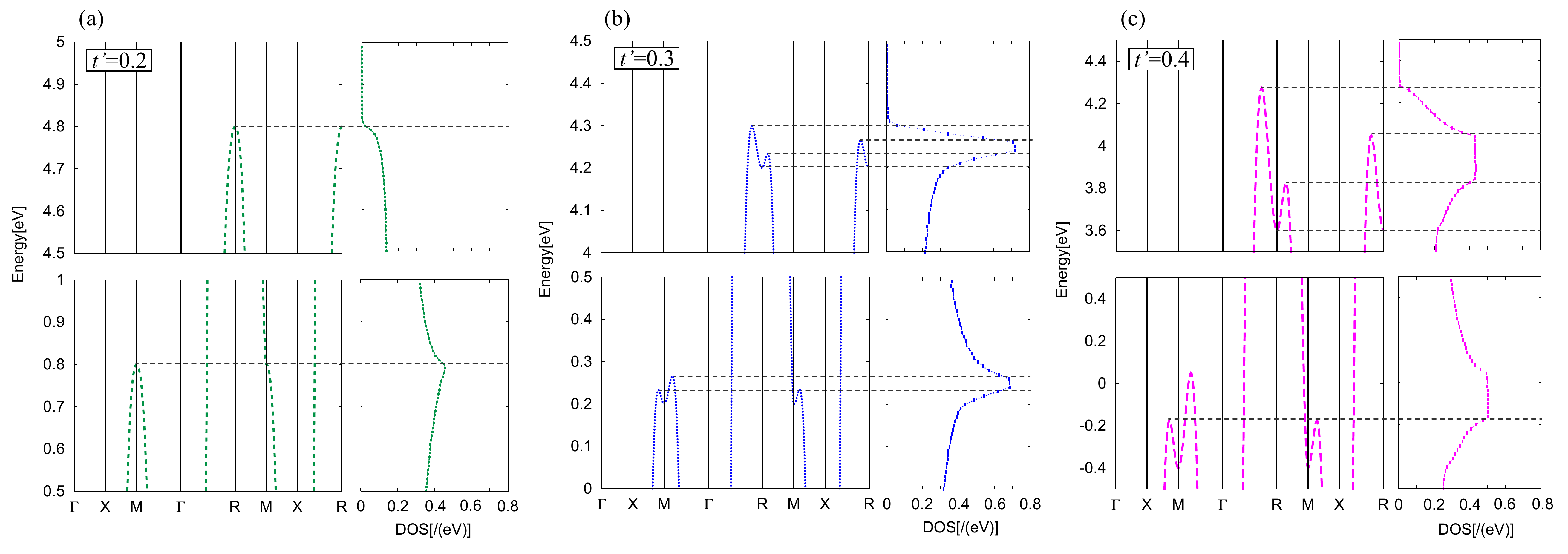}
        \caption{Close up views of the band structure and DOS around the DOS peaks for (a) $t'=0.2t$, (b) $0.3t$ and (c) $0.4t$. The dashed lines are guides to the eye. The small difference of the DOS values from those in Fig.~\ref{fig:sc-band} is due to the smearing scheme used for the DOS calculation.}
        \label{fig:sc-band-close}
    \end{center}
\end{figure*}

\begin{figure*}[t!]
    \begin{center}
        \includegraphics[scale=0.15]{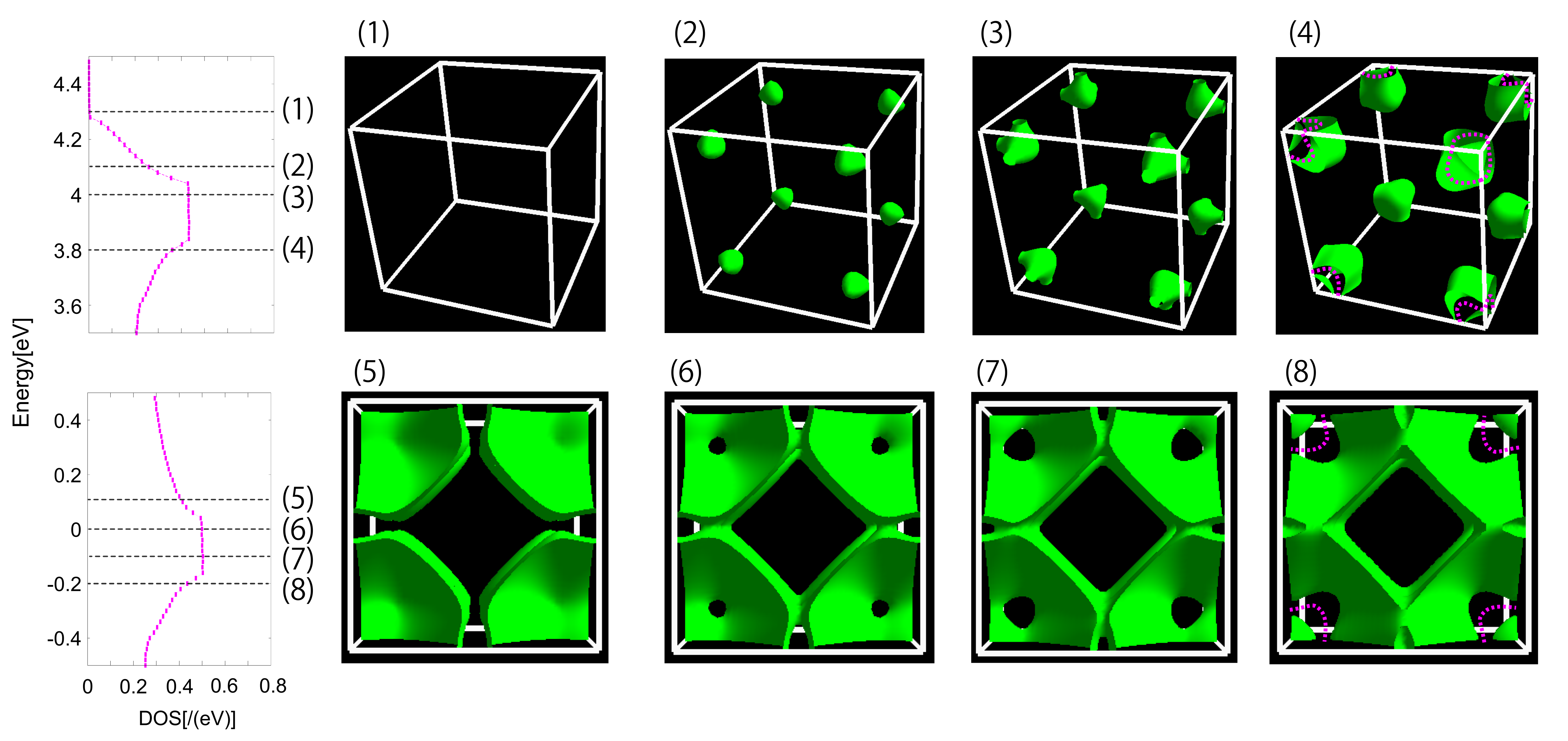}
        \caption{Snapshots of the higher order Lifshitz transitions for the equal energy surfaces with $t'=0.4t$. The simple cubic Brillouin zones are depicted as white frames. Labels (1)--(8) indicates the energy levels at which the equal energy surfaces were calculated. (upper) The surface pair formation transition around the extremum shell indicated by dashed lines. (lower) The edge-pair switching transition around the saddle loop indicated by dashed lines.}
        \label{fig:sc-LTs}
    \end{center}
\end{figure*}

\subsection{Simple cubic lattice model}
We consider the single orbital nearest neighbor tight-binding model on the simple cubic lattice [Fig.~\ref{fig:sc-band}(a)]
\begin{eqnarray}
\mathcal{H}
=-t\sum_{\langle {\bf R}, {\bf R}' \rangle_{\rm sc}} c^{\dagger}({\bf R})c({\bf R}')
,
\end{eqnarray}
where $\langle {\bf R}, {\bf R}' \rangle_{\rm sc}$ denotes that ${\bf R}$ and ${\bf R}'$ are the nearest neighbor sites in the simple cubic lattice. The resulting energy eigenvalue of the total Hamiltonian reads
\begin{eqnarray}
\varepsilon({\bf k})
&=&
-2t({\rm cos}k_{x}+{\rm cos}k_{y}+{\rm cos}k_{z})
. 
\end{eqnarray}
In the simple cubic Brillouin zone [Fig.~\ref{fig:sc-band}(b)], the isolated vHSs of types $A_{\rm 3D}$, $B_{\rm 3D}$, $B'_{\rm 3D}$ and $A'_{\rm 3D}$ are located at the $\Gamma$, $X$, $M$ and $R$ points, respectively. Therefore, the DOS does not exhibit divergence [solid lines in Fig.~\ref{fig:sc-band}(c)(d)].

In order to induce the band deformation around the critical points at the $M$ and $R$ points, we here introduce the farther neighbor hopping $t({\bf R}\!\!-\!\!{\bf R}' \!\!=\!\!2{\bf a}_{i})=-t'$ with ${\bf a}_{i} (i=x,y,z)$ being the primitive lattice vectors. The energy eigenvalue is then modified to
\begin{eqnarray}
\varepsilon({\bf k})
&=&
-2t({\rm cos}k_{x}+{\rm cos}k_{y}+{\rm cos}k_{z})
\nonumber \\
&&-2t'({\rm cos}2k_{x}+{\rm cos}2k_{y}+{\rm cos}2k_{z})
. 
\label{eq:ek-sc-tp}
\end{eqnarray}

The necessary condition for turning the saddle point at $M$ to minimum is that the maxima in the band dispersions along the $X$-$M$ and $\Gamma$-$M$ are located in the middle of the paths, which corresponds to $t' > t/4$. This is also the necessary condition for turning the maximum at $R$ to minimum. We therefore examined the band structure and DOS with $t'$ changed through $t/4$. In Fig.~\ref{fig:sc-band}(c), we indeed find maxima with $t' =0.3t$ and $0.4t$ at fractional points near $M$ and $R$, and concomitantly the peaks in the DOS are formed. In the present case, the critical points responsible for the shoulders of the DOS peaks are located exactly on the regular Brillouin zone paths (Fig.~\ref{fig:sc-band-close}). It must be, however, noted that those points are just cross sections of the high dimensional saddle loops and extremum shell, and all the states in the vicinity of those high dimensional structures contribute to the peaks. We also note that the critical points at the $M$ and $R$ points are, as displayed in Fig.~\ref{fig:sc-band-close} (b) and (c), no longer responsible for any appreciable structures in the DOS peaks.

To verify the saddle loop and extremum shell, we also examined the evolution of the equal energy surfaces through the DOS peaks with $t'=0.4t$. The surface pair formation at the extremum shell is ideally observed when the energy level is reduced through the corresponding DOS singularity. The upper panels of Fig.~\ref{fig:sc-LTs} show the change of the equal energy surfaces. The pocket formation is first observed at the maximum between the $\Gamma$-$R$ path, the surfaces gradually evolves and finally bifurcates into two nested surfaces sandwiching the deformed extremum shell (indicated by bold dashed line) when the energy level is below the lower DOS shoulder. Comparing the end points [(1) to (4)], the whole change is interpreted as the surface pair formation. Similarly, the edge pair switching transition should be observed when the energy level passes down the energy position of the ideal saddle loop. According to the lower panels, the transition starts with the neck formation at a point on the $\Gamma$-$M$ line and the resulting hole gradually evolves into the void running along the deformed saddle loop (indicated by bold dashed line), and thus the edge pair switching transition is completed. The higher-order Lifshitz transitions occurring on the single band within the nonzero energy range thus indicates the saddle loop and extremum shell.

\section{DOS peak in the cubic H$_{3}$S superconductor}
In this section, we study the electronic structure of the recently discovered high-temperature superconductor H$_{3}$S~(Refs.~\onlinecite{Eremets} and \onlinecite{Shimizu}). After its theoretical and experimental discoveries, first principle calculations have revealed its band structure.\cite{Duan2014, Bernstein-Mazin-PRB2015,Papacon-Pickett-ele-str-PRB2015,Flores-Livas2016,Quan-Pickett-vHs-PRB2016, Fan-H3AB-JPCM2016} Its particularly interesting feature is the peak in the DOS at the Fermi level. According to the Bardeen-Cooper-Schrieffer theory of the phonon-mediated conventional superconductors,\cite{BCS1, BCS2} an approximate formula of the superconducting transition temperature $T_{\rm c}$ is written as
\begin{eqnarray}
T_{\rm c}
\propto
\omega
{\rm exp}[-\frac{1}{\lambda}]
,
\end{eqnarray}
where $\omega$ denotes the frequency of the phonon mode mediating the electron pairing, and $\lambda$ is the dimensionless parameter representing the total pairing strength. Because $\lambda$ is proportional to the DOS at the Fermi level, the peaked DOS is thought to be a crucial factor for the high $T_{\rm c}$. In fact, several groups have reported that the experimentally observed $T_{\rm c}$ is accurately reproduced with the Eliashberg theory~\cite{Duan2014, Errea-PRL2015, Sano-vHS-PRB2016, Errea-Nature2016} and density functional theory for superconductors,\cite{Flores-Livas2016, Akashi-PRB2015, Akashi-Magneli-PRL2016} where its $T_{\rm c}$ is boosted by the large DOS. Although the conventional phonon mechanism thus explains the high $T_{\rm c}$, potential impacts of the DOS shoulder as the vHS and small electronic energy scale indicated by the DOS peak width have also attracted attention for possible unconventional mechanism.\cite{Bianconi-NSM, Bianconi-EPL, Bianconi-scirep} In either direction, the presence of the DOS peak at the Fermi level is a pivotal basis of the theoretical discussions. Nevertheless, its existence has yet been verified with direct experimental observations because of the difficulty of the high pressure experiments. It is hence important to see if there is any general mechanism behind the formation of the DOS peak that makes it persist against subtle changes of the structure and calculation conditions, or it is a consequence of the accidentally fine-tuned crystal structure.

The ``minimal" modeling of the electronic states in H$_{3}$S has been first proposed by Bernstein and coworkers.\cite{Bernstein-Mazin-PRB2015} They have pointed out that the inversion of the on-site energies between the hydrogen-1$s$ and sulfur 3$p$ is important and proposed a simple tight binding model with the sulfur 3$s$, 3$p$ and hydrogen 1$s$ orbitals with only the nearest neighbor hopping parameters. Later, Quan and Pickett~\cite{Quan-Pickett-vHs-PRB2016} have published a set of tight binding parameters with farther neighbor hopping obtained by the construction of the Wannier functions.\cite{MLWF1, MLWF2} Ortenzi and coworkers, on the other hand, have argued that important hopping parameters are missing in their models and proposed an alternative model~\cite{Ortenzi-TBmodel-PRB2016} constructed by the recipe of Slater and Coster.\cite{Slater-Coster} They focused on the critical point in the middle of the $N$-$X$ path in the base centered cubic (BCC) Brillouin zone and imposed a criterion that the energy level of this point, as well as the topology of the Fermi surfaces, are reproduced, but their resulting DOS peak is rather broadened compared with the first-principle calculation. Souza and Marsiglio~\cite{Souza-Marsiglio-IJMPB,Souza-Marsiglio-PRB} have took an opposite approach; they studied the features of the band structure intrinsic to the Bravais lattices. They recalled the finding by Jelitto~\cite{Jelitto1969} that the single $s$ orbital nearest neighbor tight binding model on the BCC lattice yield the divergence of the DOS, which is due to the completely flat band dispersion on planar manifolds (boundaries of the reduced simple cubic Brillouin zone) induced by the interference of the Bloch phase of the wave function.\cite{Akashi-interfere-PRB2017} However, its relation to the DOS peak of H$_{3}$S is not apparent as such flat dispersion is yet found.

For the modeling of the system, it is important to determine which of the features of the first-principle band structure are essential. We concentrate on its DOS peak; we first examine the mechanism how the DOS peak is formed and later explore the simple model that at least retains the DOS peak formed by the common mechanism. Through the previous close analysis, the DOS peak has been found to have two adjacent shoulder structures and the locations of the corresponding critical points in the ${\bf k}$ space have also been specified.~\cite{Quan-Pickett-vHs-PRB2016} We first see that the actual feature responsible for the whole peak is the above mentioned saddle loop and the previously found critical points are just its cross sections.

\begin{figure}[t]
    \begin{center}
        \includegraphics[scale=0.50]{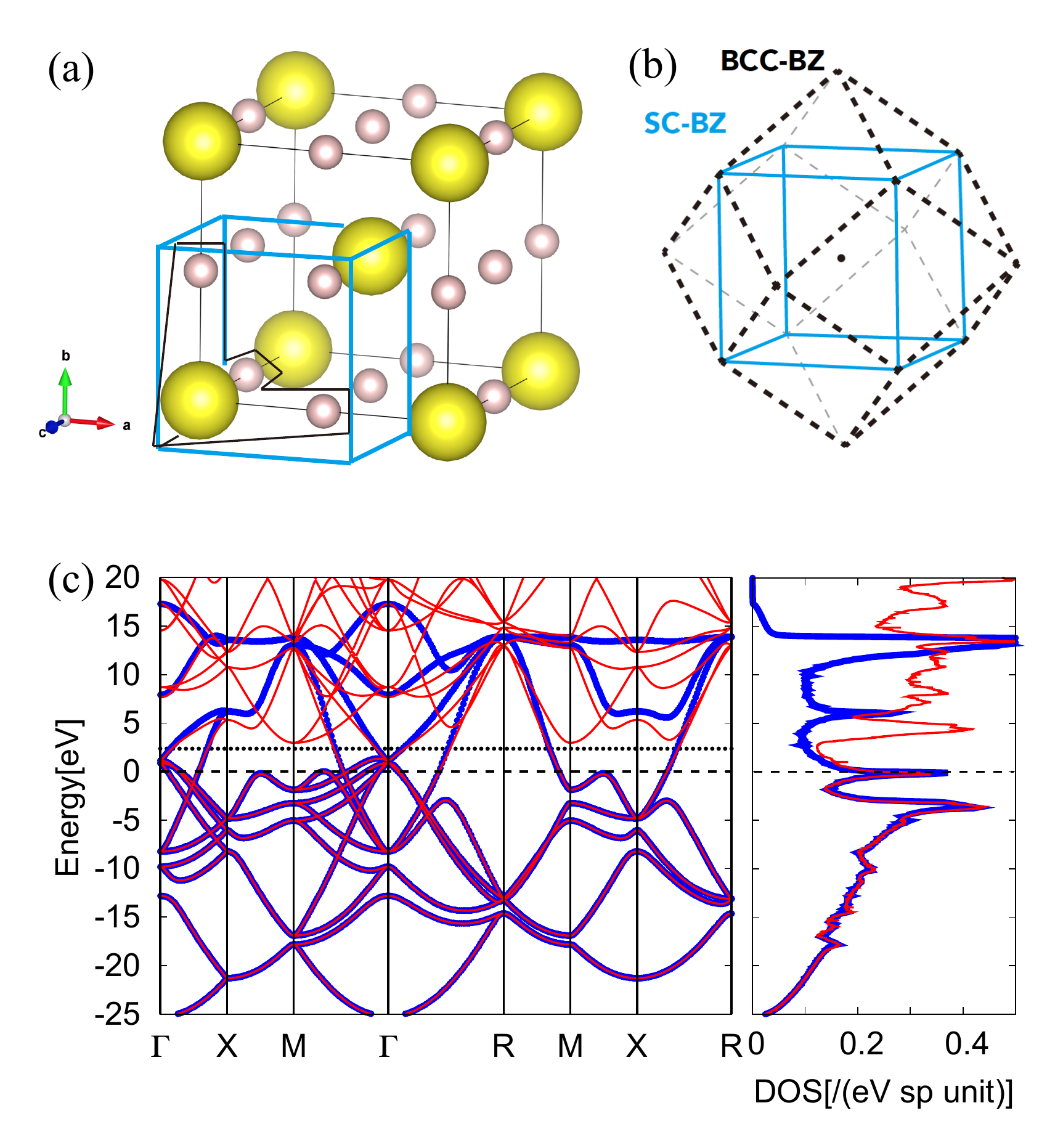}
        \caption{(a) Crystal structure of the cubic H$_{3}$S, where large and small balls represent sulfur and hydrogen atoms, respectively. The primitive and conventional unit cells are indicated by thin and bold lines, respectively. (b) The Brillouin zones corresponding to the respective definitions of the unit cell. (c) The electronic (left) band structures and (right) DOS spectra, where dashed line indicates the Fermi level. Thin and bold lines are the results of the first principle calculation and effective tight binding model derived from the Wannier functions, respectively. The dotted line is the upper bound of the frozen window for the construction of the Wannier functions (see main text).}
        \label{fig:H3S-str-band}
    \end{center}
\end{figure}

\begin{figure*}[t]
    \begin{center}
        \includegraphics[scale=0.30]{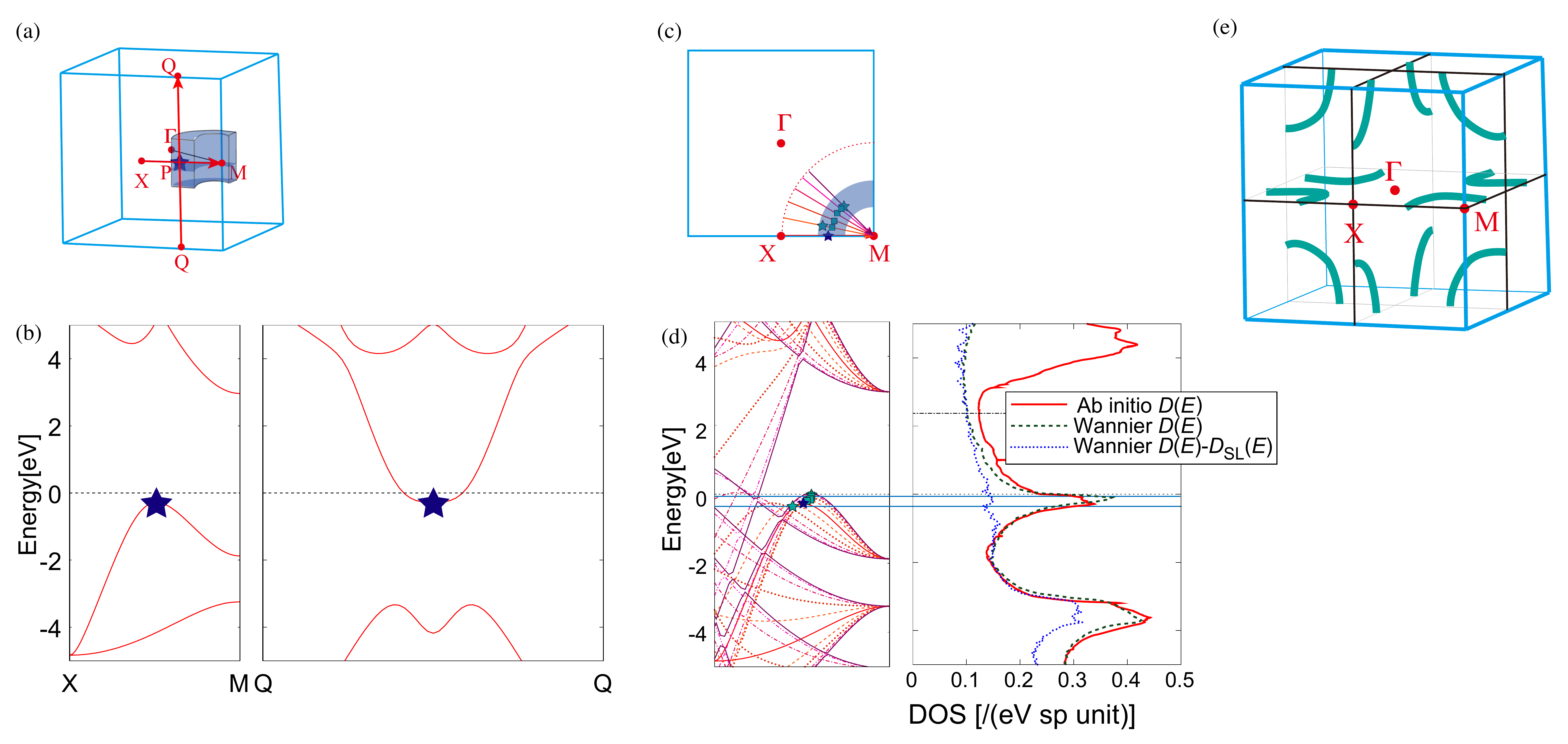}
        \caption{Band dispersions along unconventional paths. (a) The ${\bf k}$-point paths passing the target point $P$ indicated by dark star and (b) the corresponding band dispersions. (c) The ${\bf k}$-point paths for location of the saddle loop and (d) (left) the corresponding band dispersions. The light stars indicate the minimum and maximum along the line of saddle points, whereas squares represent the midpoints of the saddle line. (right) Total DOS $D(E)$ from first principles, from the Wannier model and partial DOS $D(E)-D_{\rm SL}(E)$ from the Wannier model. The region of integration for $D_{\rm SL}(E)$ ($\Omega_{\rm SL}$) [shown by shaded volume in panels (a) and (c)] is defined by $\{{\bf k} | 0.20 < \sqrt{(k_{x}-0.5)^2 + (k_{y}-0.5)^2} < 0.35; |k_{z}| < 0.15\}$ with ${\bf k}$ in the unit of $2\pi/a$. The horizontal dot-dashed line indicates the upper bound of the frozen window same as in Fig.~\ref{fig:H3S-str-band}(c). (e) Schematic picture of the saddle loops, where the anomaly is concentrated.}
        \label{fig:H3S-special-path}
    \end{center}
\end{figure*}

\subsection{First principle band structure}
We calculated the first principle electronic band structure of the cubic H$_{3}$S with the plane wave pseudopotential method as implemented in {\sc quantum espresso}.\cite{QE} The Perdew-Burke-Ernzerhof generalized gradient approximation~\cite{GGAPBE} was adopted for the exchange correlation functional. The pseudopotentials were made with the Troullier-Martins scheme.\cite{TM, FHI} The plane-wave cutoff for the wave function was set to 80~Ry. The cubic lattice parameter was set to 5.6367~bohr, which is the optimized value under 200GPa. We derived the Wannier model using {\sc wannier90}.\cite{wannier90} The sulfur-$s$, $p$ and hydrogen $s$ orbitals were adopted as the initial guess. We imposed the frozen window constraint,\cite{MLWF2} within which the Hilbert space is assured to be spanned by the resulting Wannier orbitals. Very wide energy window up to Fermi level plus $\gtrsim$ 40 eV, within which the states are used for constructing the Wannier orbitals, was set for accurate reproduction of the first principle band structure. We did not minimize the gauge-dependent part of the Wannier spreads to get the projected Wannier orbitals rather than the maximally localized ones so that subtle possible shift of the Wannier centers from the atomic sites does not occur.~\cite{PhysRevB.74.195118}

We show the calculated band structure in Fig.~\ref{fig:H3S-str-band} (c). Here we took the non-primitive simple cubic cell~\cite{Bianconi-scirep} [panel (a)] for convenience in the later discussions, by which the Brillouin zone is folded from the BCC one as shown in panel (b). Our calculated band structure (thin line) reproduces the features shown in the previous studies, especially the sharp peak in the DOS at the Fermi level.\cite{Quan-Pickett-vHs-PRB2016} The apparent local maxima seen in the middle of the $X$--$M$ and $M$--$\Gamma$ paths have been related to the DOS peak in previous discussions.\cite{Bianconi-scirep} We name the maximum along the $X$--$M$ path $P$ for later analysis. We further scrutinize the relation between the details of the band structure and DOS peak.

\begin{figure}[t]
    \begin{center}
        \includegraphics[scale=0.50]{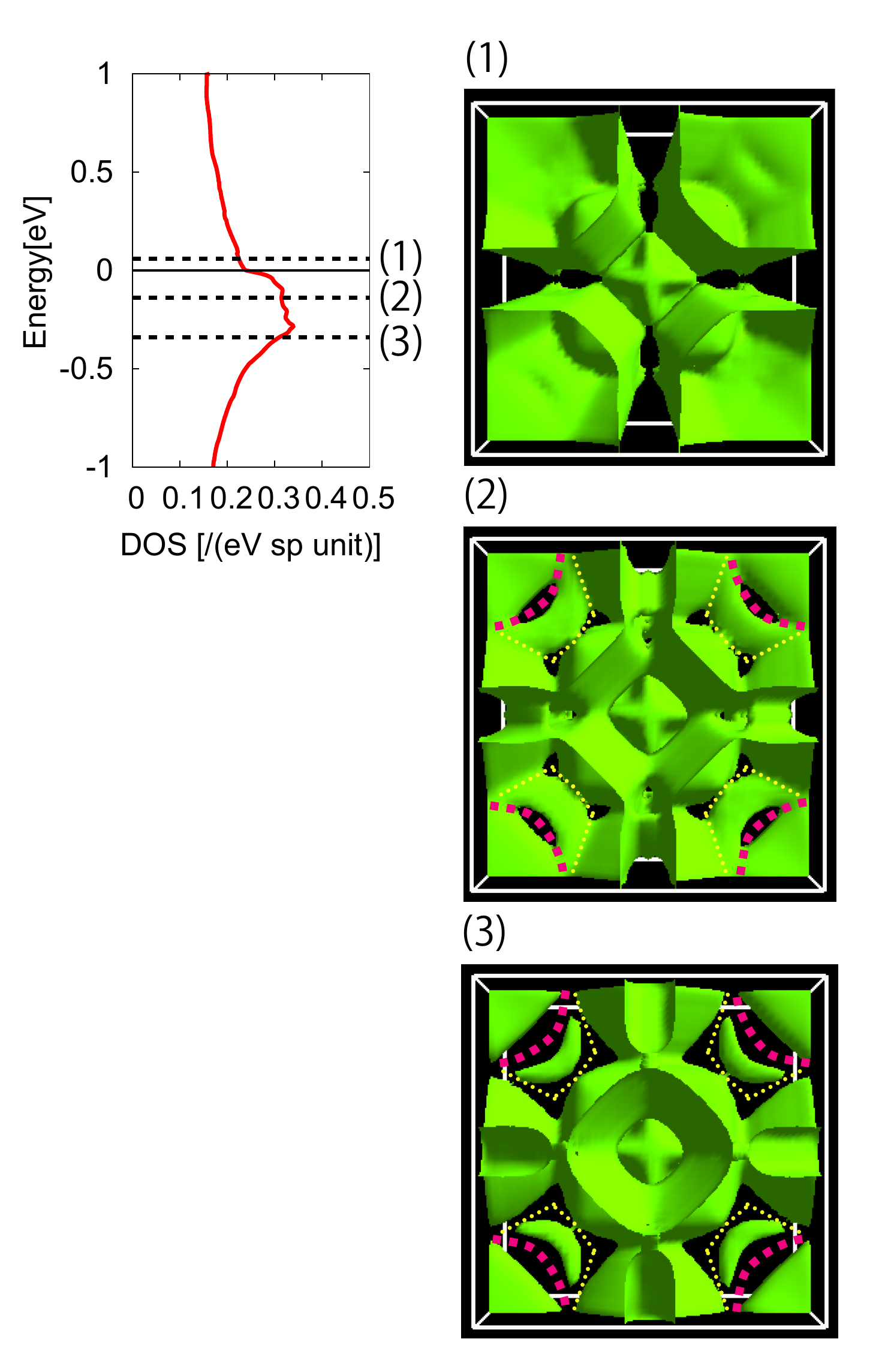}
        \caption{Edge switching transition in the equal energy surfaces with variable energy levels through the DOS peak in H$_{3}$S. The bold and thin dashed lines indicate the transition lines which are and are not responsible for the DOS peaks, respectively.}
        \label{fig:H3S-edge-switch}
    \end{center}
\end{figure}

We find that the band dispersion along $k_{z}$ through $P$ is convex as shown in Fig.~\ref{fig:H3S-special-path} (b). To see the band dispersion in the direction normal to the BZ boundary, we also calculated the band dispersions along the paths depicted in Fig.~\ref{fig:H3S-special-path} (b). Departing from the boundary, the degenerate bands split, and one of the maxima are aligned in a curly line indicated by stars. Those points, around which the band is concave in the path directions and convex in the $k_{z}$ direction, form the saddle loop encircling the $M$ point. Note that there are in total three symmetrically equivalent saddle loops interchanged by the three fold rotation in the (1 1 1) axis. We find that the small dispersion along the loop corresponds to the width of the DOS peak [panel (d)], which suggests that the DOS peak is dominated by the saddle loops. Furthermore, we calculated the local DOS originating from the saddle loops by
\begin{eqnarray}
D_{\rm SL}(E)=N_{\rm sym}\sum_{n{\bf k}\in \Omega_{\rm SL}}\delta(E-\varepsilon_{n{\bf k}})
,
\label{eq:D-SL}
\end{eqnarray}
where the volume of integration $\Omega_{\rm SL}$ is limited to a tiny shaded region depicted in Fig.~\ref{fig:H3S-special-path}(a)(c) and $N_{\rm sym}$(=12) is the symmetry factor. By subtracting $D_{\rm SL}(E)$ from the total DOS $D(E)$, the peak structure completely vanished [Fig.~\ref{fig:H3S-special-path}(d)]. This result directly evidences that the DOS peak structure is wholly due to the saddle loop. 

Finally, we also calculated the equal energy surfaces with different energy levels through the DOS peak position. As indicated by dashed lines in Fig.~\ref{fig:H3S-edge-switch}, we observed two edge pair switching transitions on the paths around the $M$ point. According to the above analysis, the region where the inner transition occurs is responsible for the DOS peak. The position of the higher order Lifshitz transition through the energy range of the target DOS peak thus indicates the ``hot spot" corresponding to the peak. %Note that the analysis of the equal energy surface for this purpose must be done band by band; the higher order Lifshitz transitions which does not relate to the DOS enhancement can appear by band crossing. 

We thus establish the simplistic view on the complicated electronic structure of this system as depicted in Fig.~\ref{fig:H3S-special-path} (e). Its anomalous aspects, peaked concentration of the DOS and possible competition of the electronic and phononic energy scales, originates from tiny regions in the ${\bf k}$ space around the $M$ points where the saddle loops are located. This fact should encourage further scrutiny on the electronic states of the saddle loop regions and their interplay with phonons, not only the multiple hole pockets around the $\Gamma$ point,~\cite{PhysRevB.92.205125,Ghosh-H3S-ph-break} though it is out of the scope of the present work.
 
\begin{figure}[t]
    \begin{center}
        \includegraphics[scale=0.50]{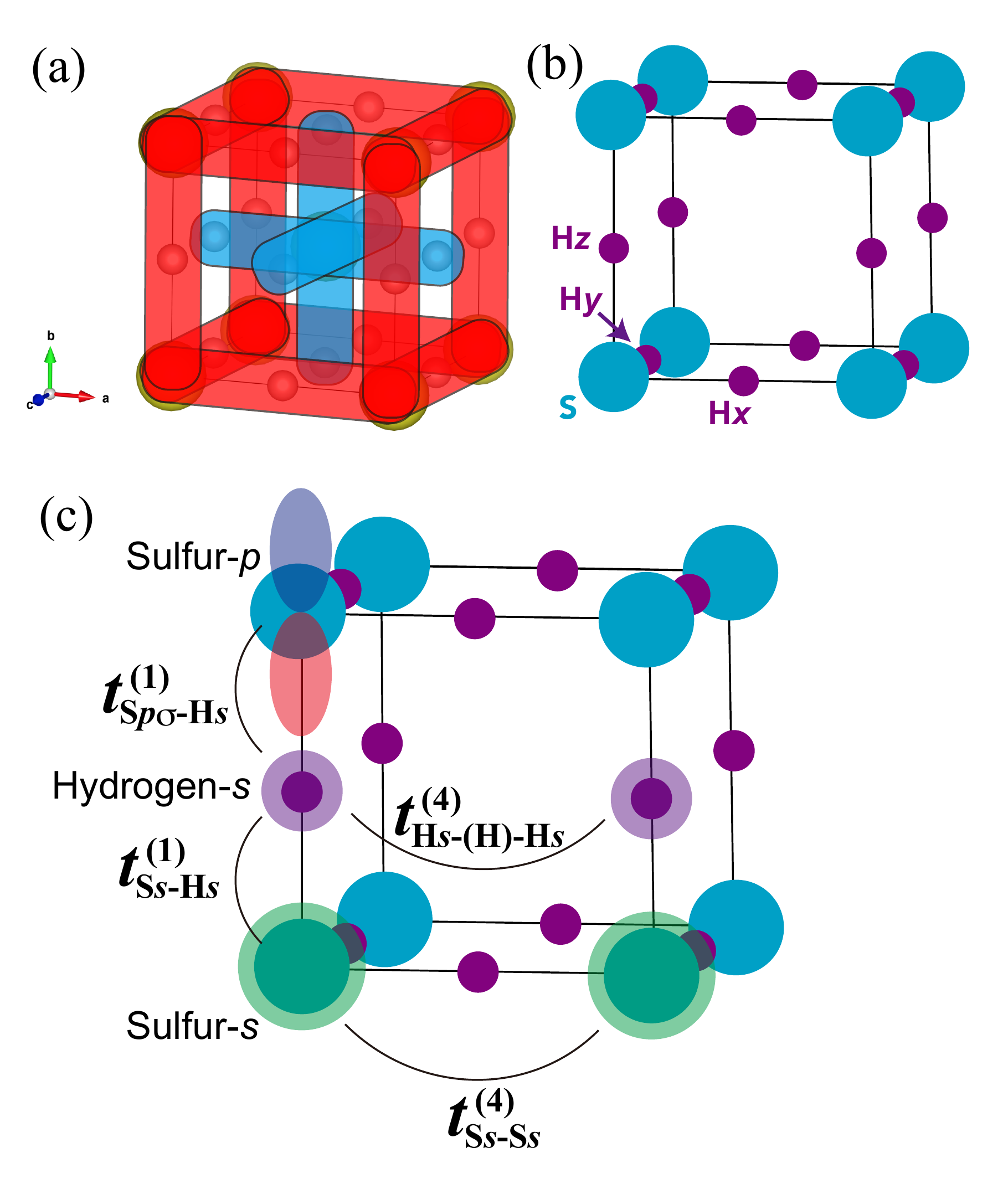}
        \caption{(a) Partitioning of the crystal structure of H$_{3}$S into two identical simple cubic lattices. (b) The simple cubic (ReO$_{3}$) H$_{3}$S structure. (c) The tight binding hopping parameters of the minimal model. The symmetrically equivalent parameters are not shown. Note that the actual form of the Wannier orbitals are depicted in Fig.~\ref{fig:Wannier}.}
        \label{fig:bipartite}
    \end{center}
\end{figure}

\subsection{Wannier model and its analysis}
\begin{figure*}[t]
    \begin{center}
        \includegraphics[scale=0.45]{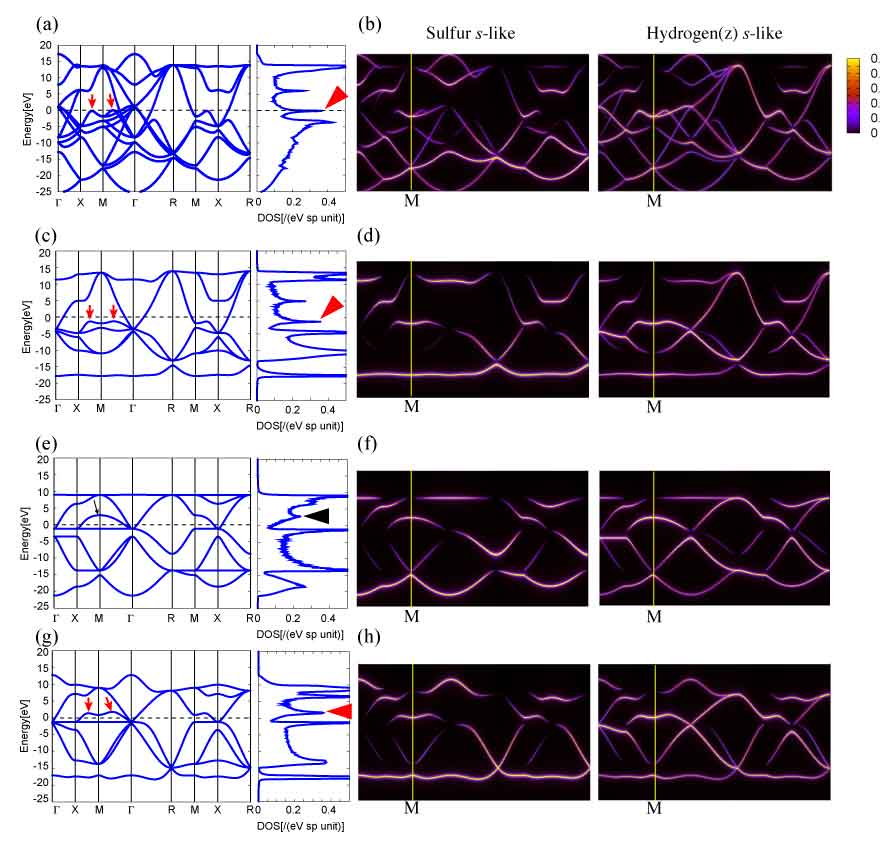}
        \caption{Electronic band structure of the Wannier models with different levels of approximation. (left) Band structures and DOS spectra derived from (a) the Wannier model that shows the perfect agreement with the first principle calculation for the low energy states [See Fig.~\ref{fig:H3S-str-band}(c)], (c) the Wannier model where the hopping between the two bipartite lattices are ignored, (e) the bipartite simple cubic Wannier model with only the nearest neighbor hopping parameters [$t^{(1)}_{{\rm S}s-{\rm H}s}$ and $t^{(1)}_{{\rm S}p\sigma-{\rm H}s}$ in Fig.~\ref{fig:bipartite}], and (d) the ``minimum" model where the farther neighbor hopping parameters [$t^{(4)}_{{\rm S}s-{\rm S}s}$ and $t^{(4)}_{{\rm H}s-({\rm H})-{\rm H}s}$ in Fig.~\ref{fig:bipartite}] are added on top of (c). (b) (d) (f) (h) The corresponding spectral functions. }
        \label{fig:Spectr}
    \end{center}
\end{figure*}

The present result motivates us to seek for a simple model that reproduces the saddle loop as the minimal model for the DOS peaking in H$_{3}$S. Starting from the seven orbital Wannier model (sulfur-$s$, -$p$ and hydrogen $s\times 3$) that perfectly reproduces the {\it ab initio} band structure [Fig.~\ref{fig:H3S-str-band}(c); see Table~\ref{tab:tbparam-app} for the original hopping parameters], we conducted a thorough examination on how the band structure and DOS change by omitting any hopping parameters. The crystal structure of the cubic H$_{3}$S is bipartite, in that it is formed by identical simple cubic sublattices shifted from each other by $(a/2, a/2, a/2)$ [Fig.~\ref{fig:bipartite}(a)]. In preceding analyses,\cite{Heil-Boeri2015, Quan-Pickett-vHs-PRB2016} it has been pointed out that the electronic density contributed to by the states near the DOS peak is spatially concentrated along the individual simple cubic frames. Inspired by this fact, we calculated the electronic structure of a ``bi-partitioned" simple cubic lattice (known as the ReO$_{3}$ structure) by neglecting all the hopping parameters across the sublattices. Interestingly, we find that the saddle loop structure around the $M$ point, as well as the DOS peak emerging from this, are retained with this simple cubic lattice model as shown in Fig. ~\ref{fig:Spectr}(c). The local maxima along the $X$-$M$ and $M$-$\Gamma$ paths indicated by arrows are the cross sections of the saddle loop around the $M$ point. 

The above result suggests an interesting possibility: the saddle loop structure is persistent against the doubling of the simple cubic cell, and then the seven-orbital simple cubic lattice model captures the mechanism of the DOS peak formation. The former point can be supported theoretically, by considering the interference effect of the Bloch functions.\cite{Akashi-interfere-PRB2017} Consider any Bloch state composed of orbitals on a single sublattice $|\Phi_{s; n{\bf k}}\rangle$, where $s(=1,2)$ denotes the index of the sublattice. The Wannier Hamiltonian $\mathcal{H}$ is block diagonalized for each ${\bf k}$. Due to the commensurate shift between the sublattices and mirror symmetries, the following formula holds
\begin{eqnarray}
\langle \Phi_{1; n{\bf k}}|\mathcal{H}|\Phi_{2; n{\bf k}}\rangle=0
\ \ ({\bf k}\in {\rm boundary\ of\ the\ SC\ BZ}).
\nonumber \\
\label{eq:sub-decouple}
\end{eqnarray}
This implies that the band structure at the BZ boundary is less affected by the hopping across the sublattices, compared with the ${\bf k}$-point regions far from the boundary. With the saddle points along the $X$-$M$ and $X'$-$M$ paths [Fig.~\ref{fig:sc-band}(b)] retained against the intersublattice coupling, from the physical requirement that the variation of the band dispersion is not drastic in the ${\bf k}$ space, the saddle loop connecting those points must be persistent as well. We hence assert that the decoupled simple cubic lattice model is the minimal one.

To seek the origin of the saddle loop, we next calculated the electronic structure of the single sublattice model with only the nearest neighbor hopping parameters retained [Fig.~\ref{fig:Spectr}(e)]. The saddle loop structure then converged to the saddle {\it point} at the $M$ point as indicated by an arrow. The corresponding DOS peak was smeared. We found that the saddle loop is recovered by considering farther neighbor hopping parameters $t^{(4)}_{{\rm S}s-{\rm S}s}$ and $t^{(4)}_{{\rm H}s-({\rm H})-{\rm H}s}$ [Fig.~\ref{fig:bipartite}(c)]. The superscript $(4)$ implies that the hopping connects the fourth nearest neighbor sites in the original crystal structure. Notably, those parameters are positive and yield contributions to the energy spectrum in the $+[{\rm cos}({\bf a}_{1}\cdot {\bf k}) +{\rm cos}({\bf a}_{2}\cdot {\bf k})]$ form, with which the energy eigenvalue at $M$ is reduced (pushed). The orbital resolved spectral function (Ref.~\onlinecite{Fetter-Walecka}; see Appendix~\ref{app:spectr} for its formal definition) displayed in Fig.~\ref{fig:Spectr}(f) shows that the electronic state at the saddle point $M$ is mainly composed of the sulfur $s$ and hydrogen H$_{z}$ $s$ orbitals [Fig.~\ref{fig:Spectr}(f)]. Introduction of $t^{(4)}_{{\rm S}s-{\rm S}s}$ and $t^{(4)}_{{\rm H}s-({\rm H})-{\rm H}s}$ therefore substantially reduces the energy eigenvalue at $M$, forming the saddle loop around it.

\begin{figure}[t]
    \begin{center}
        \includegraphics[scale=0.50]{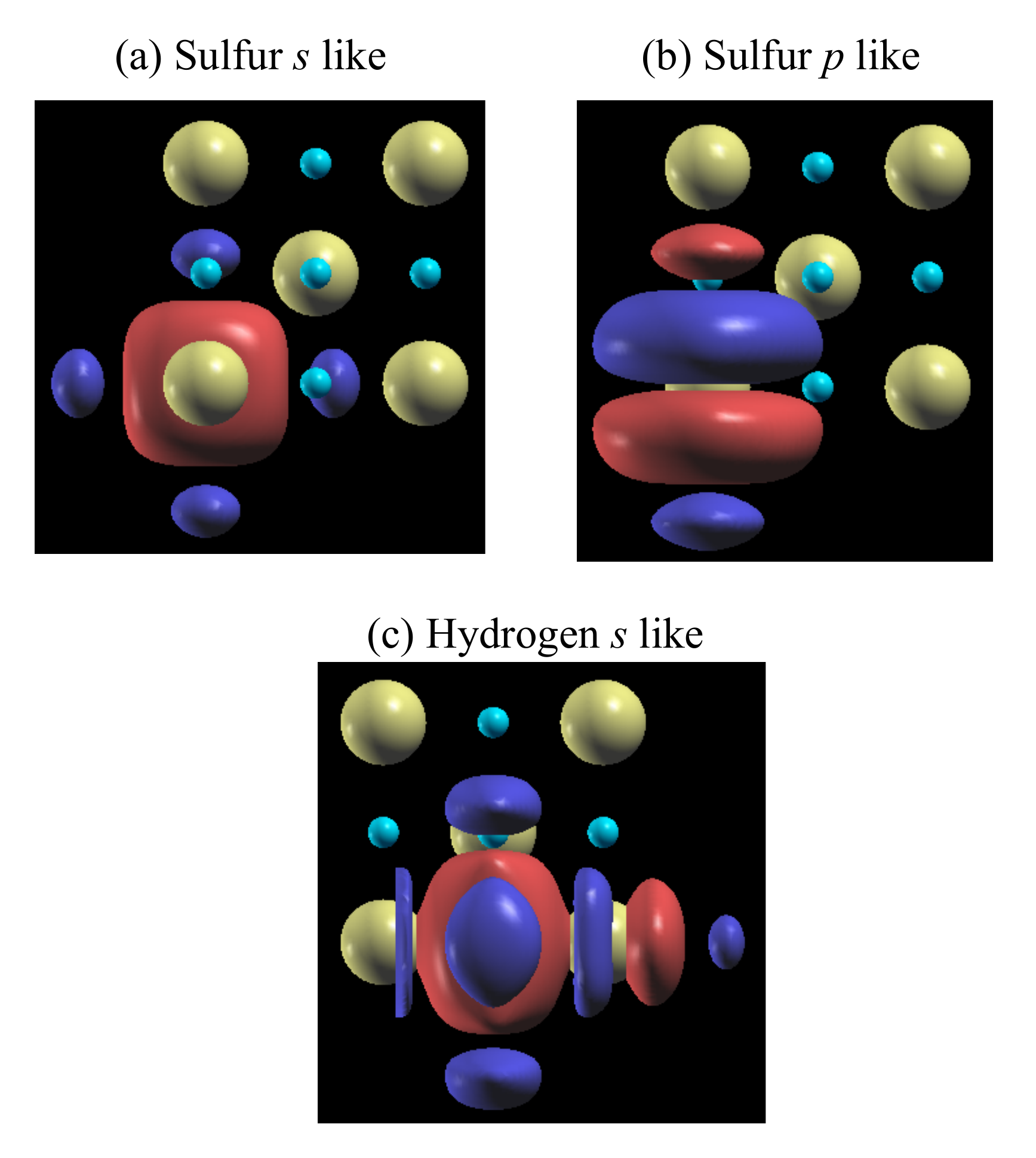}
        \caption{Isosurfaces of the calculated Wannier functions, where signs of the functions are distinguished by colors; (a) Sulfur $s$-like orbital, (b) sulfur $p$-like orbital and (c) hydrogen $s$-like orbital.}
        \label{fig:Wannier}
    \end{center}
\end{figure}

The origin of the large farther neighbor hopping with positive sign can be suggested from the Wannier orbitals displayed in Fig.~\ref{fig:Wannier}. Usually, the hopping between two orbitals without phase shift is negative. In the present case, however, both the sulfur $s$-like and hydrogen $s$-like orbitals show the sign inversion of the wave functions and have tails in a length scale comparable to half of the lattice parameter. These features cause the relatively large farther neighbor hopping with the sign inversion.

\begin{table*}[t!]
\caption{Tight binding parameters (eV) of the minimal model that explains the saddle loop around the $M$ point and related DOS peak. The onsite energies are calculated from the level of sulfur $s$-like orbital. The sign $\pm$ indicates the arbitrariness in the definition of the $s$-$p\sigma$ type hopping. See Fig.~\ref{fig:bipartite} (c) for the definitions of the hopping parameters.}
\begin{tabular}{|c|cccc|cc|}
 \hline
 Pressure (GPa) &120 &160& 200&240&\multicolumn{2}{c|}{120} \\ \hline
 Distortion &\multicolumn{4}{c|}{No}& Molecule & Lattice \\ \hline 
 Lattice parameter (a.u.) &5.8990& 5.7542&5.6367&5.5375  &\multicolumn{2}{c|}{(see text)}\\ \hline
  Onsite &&&& &&\\
  Sulfur $s$ like &---&---& --- & ---&---& ---\\
  Sulfur $p$ like &8.16&7.82& 7.49  & 7.20 &7.98&8.32\\
  Hydrogen $s$ like &6.42&5.77& 5.18  & 4.62 &6.27&6.50\\ \hline
  Hopping & &&&&&\\
  $t^{(1)}_{{\rm S}s-{\rm H}s}$ &-3.84&-4.09& -4.30 & -4.48 &-4.69, -2.98&-3.84\\
  $t^{(1)}_{{\rm S}p\sigma-{\rm H}s}$ &$\pm$5.04&$\pm$5.35& $\pm$5.62  & $\pm$5.85 &$\pm$5.69, $\pm$4.31&$\pm$5.07\\
  $t^{(4)}_{{\rm S}s-{\rm S}s}$ &+0.81&+0.91& +0.99 & +1.07 &+0.78&+0.81 \\
  $t^{(4)}_{{\rm H}s-({\rm H})-{\rm H}s}$ &+0.43&+0.46& +0.49  & +0.51 &+0.45&+0.46\\ \hline
\end{tabular}
\label{tab:tbparam}
\end{table*}

\begin{figure*}[t]
    \begin{center}
        \includegraphics[scale=0.50]{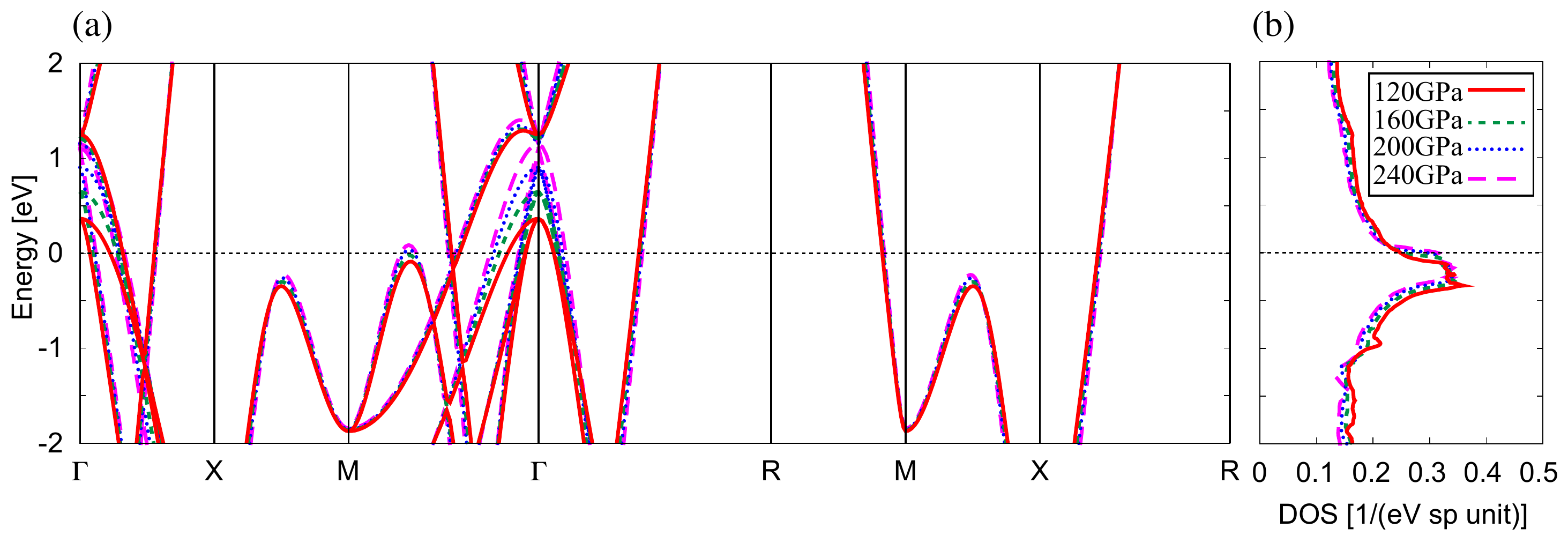}
        \caption{Pressure dependence of the first principles (a) band structure and (b) DOS for the cubic H$_{3}$S. The scale of the {\bf k}-point axis, which actually depends on the lattice parameter, is fixed for comparison.}
        \label{fig:pressure-band-dos}
    \end{center}
\end{figure*}

\subsection{Summary: what makes the DOS peaked?}
We have thus reached the minimal model of H$_{3}$S with seven (or six, if we regard the onsite energy of the Sulfur-$s$ like orbital arbitrary) parameters (Table~\ref{tab:tbparam}) that demonstrates the DOS peak formation, The basis set is composed of the sulfur $s$, $p$ and hydrogen $s$ orbitals with structural deformation in the cubic ReO$_{3}$ configuration. The nearest neighbor hopping model gives rise to a band with small dispersion, on which a saddle point is located at $M$. The further nearest neighbor hopping modifies this saddle point into the local minimum and the saddle loop is formed around it. The doubling of the lattice substantially modifies the whole electronic structure, but its important features---the maxima along the $X$-$M$ and $X'$-$M$ paths---are protected due to the interference of the Bloch states, which forces the saddle loop through those points to persist. The relevance of the present model to the actual H$_{3}$S is validated by analysis of the spectral functions [Figs.~\ref{fig:Spectr} and \ref{fig:Spectr-app}]. The {\bf k} point dependence of the orbital character around the saddle loop is surprisingly consistent among the four models; the original model that perfectly reproduces the {\it ab initio} band structure, bipartite one, bipartite nearest neighbor one and the final minimal one. 

In the above description we ignored the hopping across the different sublattices in order to highlight the mechanism that ``pushes" the saddle critical point. We also note a quantitative effect of some intersublattice hopping parameters. We found that concurrent inclusion of $t^{(1)}_{{\rm H}s-{\rm H}s}$ and $t^{(2)}_{{\rm S}p\sigma-{\rm H}s}$ (Table~\ref{tab:tbparam-app}) in combination with the farther neighbor intrasublattice hopping parameters $t^{(4)}_{{\rm S}s-{\rm S}s}$ and $t^{(4)}_{{\rm H}s-({\rm H})-{\rm H}s}$ have also effect of enhancing the ``M"-shape dispersion of the relevant band along the $X-M-\Gamma$ path in Fig.~\ref{fig:Spectr}(c), though its mechanism is not obvious. The original first principles band structure is then reproduced better. Note that those parameters do not solely yield the saddle loop, whereby we conclude that their effect is secondary compared with the farther neighbor hopping. The detail on this point is summarized in Appendix~\ref{app:intersub}.

\begin{figure*}[t]
    \begin{center}
        \includegraphics[scale=0.40]{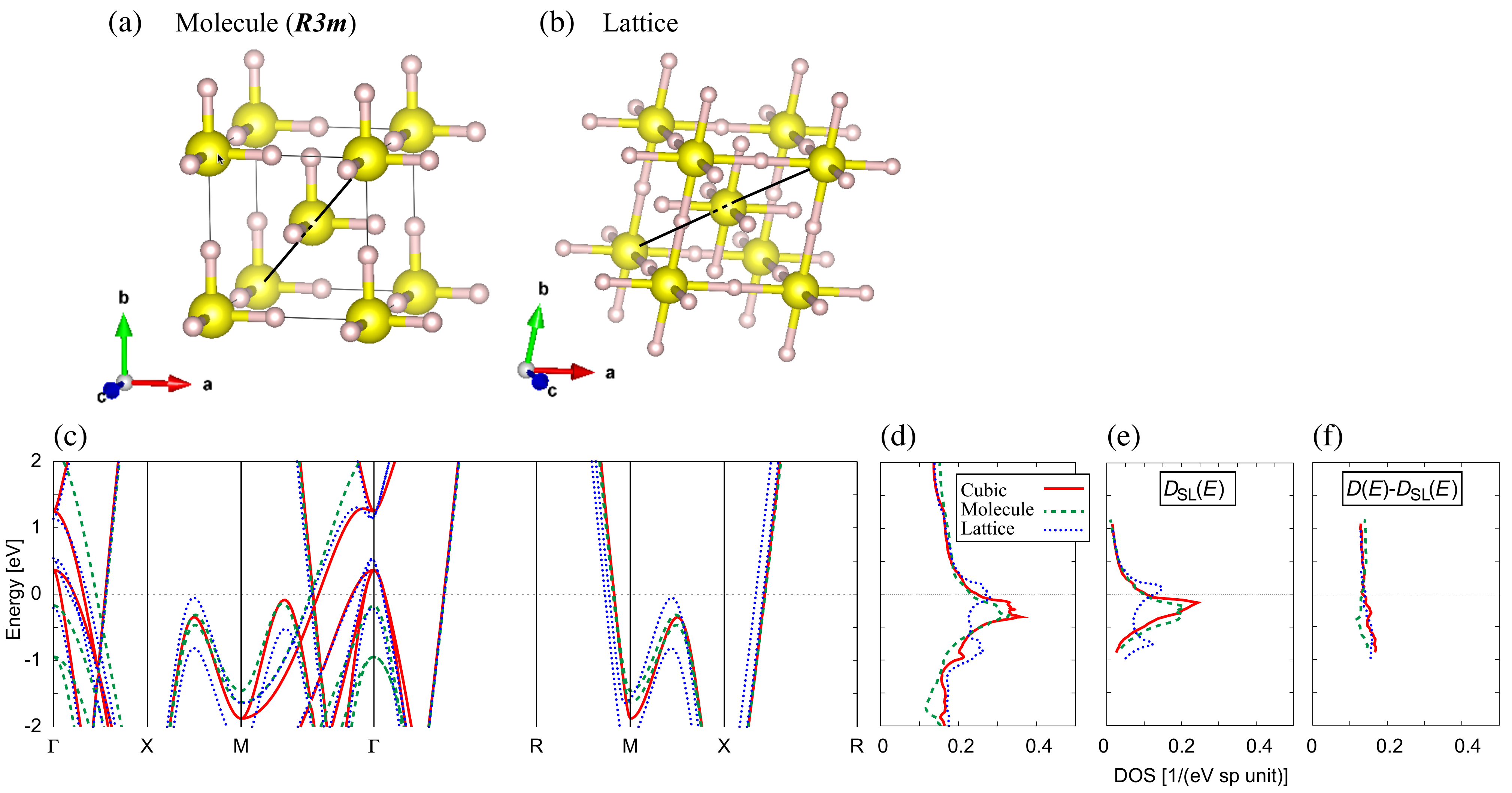}
        \caption{Distortion dependence of the band structure and DOS at 120GPa. Although the reciprocal vectors vary by the distortions, we plot the bands in the same scale by defining the special points in terms of the reciprocal vector coordinates. (a)(b) Views of the distortions, where the (1 1 1) axis is indicated by solid line. The degrees of the distortions are exaggerated. (c) and (d) are the first principles band structures and DOS with and without the distortions, respectively. (e) The saddle loop contribution $D_{\rm SL}(E)$ defined in Eq.~(\ref{eq:D-SL}) and (f) the remnant $D(E)-D_{\rm SL}(E)$, calculated with the Wannier models.}
        \label{fig:distort-band-dos}
    \end{center}
\end{figure*}

Finally, we try to reconcile the previous representative proposals of the tight binding modeling of this system [Refs.~\onlinecite{Bernstein-Mazin-PRB2015,Quan-Pickett-vHs-PRB2016,Ortenzi-TBmodel-PRB2016}] with our ``minimal" explanation in view of the reproduction of the saddle loops and DOS peaks. Among them, only Ref.~\onlinecite{Ortenzi-TBmodel-PRB2016} employed a different method for calculating the parameters; prepare the model first and fit the parameters so that the band energies at selected {\bf k} points agree with the first-principles values. In Table~\ref{tab:tbparam-app}, we summarize the first, second, third and fourth neighbor hoppings connecting the sites departed by distances $a/2$, $a/\sqrt{2}$, $\sqrt{3}a/2$ and $a$, respectively. The subscripts ${\rm S}s$, ${\rm S}p$, and ${\rm H}s$ denote the Wannier orbitals displayed in Fig.~\ref{fig:Wannier}. To our observation, the fifth neighbor hopping parameters (across the sites departed by $\sqrt{5}a/2$) also have appreciable effects on the whole band structure, which we do not append in the text, though. The full set of the parameters is available online in the output format of {\sc Wannier90}.\cite{Wannier-github}

 The inversion of the onsite energy levels of the sulfur $p$(-like) and hydrogen $s$(-like) states are wholly consistent among all the models. Also, the {\it published} Wannier model parameters in Refs.\onlinecite{Bernstein-Mazin-PRB2015,Quan-Pickett-vHs-PRB2016} are in fair agreement with ours, where the subtle differences are thought to originate from the gauge degree of freedom of the Wannier orbitals or setting of the energy window parameters. Bernstein and coworkers~\cite{Bernstein-Mazin-PRB2015} published only the nearest neighbor hopping parameters; we found that those can yield DOS peak related to isolated saddle points at $M$ similarly to Fig.~\ref{fig:Spectr} (e) but not the saddle loop. The list published by Quan and Pickett~\cite{Quan-Pickett-vHs-PRB2016} includes the positive farther neighbor hopping parameters ($t^{(4)}_{{\rm S}s-{\rm S}s}$ and $t^{(4)}_{{\rm H}s-({\rm H})-{\rm H}s}$ in our expression), from which we expect that the DOS peak as well as the saddle loops are sufficiently reproduced. They also reported that the DOS calculated with the single sublattice model with the nearest neighbor hopping is ``nothing like the original one". This result probably corresponds to our Fig.~\ref{fig:Spectr} (e); incorporating $t^{(4)}_{{\rm S}s-{\rm S}s}$ and $t^{(4)}_{{\rm H}s-({\rm H})-{\rm H}s}$ into their sublattice model would reproduce the DOS peak and saddle loops. The hopping parameter between sulfur $s$-like and $p$-like states departed by $(a/2,a/2,a/2)$ ($W_{sp}$ in Ref.~\onlinecite{Ortenzi-TBmodel-PRB2016}; $t^{(3)}_{{\rm S}s-{\rm S}p\sigma}$ in our expression) was, in our calculation, small, which is probably why Quan and Pickett did not mention this parameter. More on this parameter, we found that it has an effect of {\it raising} the energy level at the midpoint of the $M$--$\Gamma$ path, which corresponds to the maximum in the $N$--$X$ path of the BCC Brillouin zone. Ortenzi and coworkers~\cite{Ortenzi-TBmodel-PRB2016} referred to this point, not the loop, for their parameter fitting and did not introduce the farther neighbor hoppings; therefore, the isolated vHS was reproduced, but the whole saddle loop structures was not, which is likely the reason of their DOS peak smearing.

\begin{table*}[t!]
\caption{Tight binding parameters of the present work and Refs.~\onlinecite{Bernstein-Mazin-PRB2015,Quan-Pickett-vHs-PRB2016,Ortenzi-TBmodel-PRB2016}. The sign $\pm$ indicates the arbitrariness in the definition of the $s$-$p\sigma$ type hopping. Dagger ($\dagger$) indicates the parameters included in our minimal model (Table~\ref{tab:tbparam}).}
\begin{tabular}{c|ccc|c}
\hline
 Parameter (eV)& Present & Ref.~\onlinecite{Bernstein-Mazin-PRB2015} & Ref.~\onlinecite{Quan-Pickett-vHs-PRB2016} & Ref.~\onlinecite{Ortenzi-TBmodel-PRB2016}   \\ \hline
 Lattice parameter (a.u.) &5.6367 & 5.6409& 5.6& 5.64 \\ \hline
 Method &\multicolumn{3}{c|}{Wannier orbital construction}& Analytical fit \\ \hline
  Onsite& & & & \\
  Sulfur $s$ like & --- & --- & --- &---\\
  Sulfur $p$ like & 7.49 & 7.3 & 7.95& 11.38\\
  Hydrogen $s$ like & 5.18 & 3.6& 2.52 &10.29\\ \hline
  1st NN & & & &\\ 
  $(\dagger)$ $t^{(1)}_{{\rm S}s-{\rm H}s}$ & -4.30 & -4.2 &-4.37 &+2.81\\
  $(\dagger)$ $t^{(1)}_{{\rm S}p\sigma-{\rm H}s}$ & $\pm$5.62 & -5.2&-5.42 &+4.65\\
   $t^{(1)}_{{\rm H}s-{\rm H}s}$ & -2.26 & -2.7 & -2.80 &-2.73\\  \hline 
   2nd NN & & & &\\
  $t^{(2)}_{{\rm S}s-{\rm H}s}$ & -0.10 & (N. A.) & (N. A.)& --- \\
  $t^{(2)}_{{\rm S}p\sigma-{\rm H}s}$ & $\pm$1.00 &(N. A.) & +0.93& --- \\ 
  $t^{(2)}_{{\rm H}s-{\rm H}s}$ & +0.08 & (N. A.) & (N. A.) & --- \\ \hline
   3rd NN & & & &\\
  $t^{(3)}_{{\rm S}s-{\rm S}s}$ & +0.06 & (N. A.)& +0.30&+2.31\\
  $t^{(3)}_{{\rm S}s-{\rm S}p \sigma}$ & $\pm$0.30 & (N. A.) & (N. A.) &+3.33 \\ 
  $t^{(3)}_{{\rm S}p-{\rm S}p \sigma}$ & +1.38 & (N. A.) &(N. A.) &+1.69\\ 
  $t^{(3)}_{{\rm S}p-{\rm S}p \pi}$ & +0.29 & (N. A.) &(N. A.) &-0.07\\ 
  $\frac{1}{4}t^{(3)}_{{\rm S}p-{\rm S}p \sigma}+\frac{3}{4}t^{(3)}_{{\rm S}p-{\rm S}p \pi}$ & +0.56 & (N. A.) &+0.60 &+0.37\\ 
  $t^{(3)}_{{\rm H}s-{\rm H}s}$ & +0.01 & (N. A.) &(N. A.) & ---\\ \hline    
   4th NN & & & &\\
  $(\dagger)$ $t^{(4)}_{{\rm S}s-{\rm S}s}$ & +0.99 & (N. A.) & +0.94&--- \\
  $t^{(4)}_{{\rm S}s-{\rm S}p \sigma}$ & $\pm$1.11 & (N. A.) & 1.29 &--- \\ 
  $t^{(4)}_{{\rm S}p-{\rm S}p \sigma}$ & -1.78 & (N. A.) & -1.83& ---\\ 
  $t^{(4)}_{{\rm S}p-{\rm S}p \pi}$ & -0.05 & (N. A.) & (N. A.)& ---\\ 
  $(\dagger)$ $t^{(4)}_{{\rm H}s-({\rm H})-{\rm H}s}$ & +0.49 & (N. A.) & +0.55& ---\\     
  $t^{(4)}_{{\rm H}s-({\rm S})-{\rm H}s}$ & -1.28 & (N. A.)& -1.14& ---\\ \hline    
\end{tabular}
\label{tab:tbparam-app}
\end{table*}

\subsection{Pressure dependence}
The two pivotal statements established through this section are that (i) the DOS peak at the Fermi level originates from the saddle loops around the $M$ and symmetrically related points and (ii) the single sublattice model explains how they are formed. We examine if those points are also valid against the change of the external pressure. We recalculated the electronic structures with the cubic lattice paremeters optimized at 120, 160, and 240 GPa and derived the Wannier model parameters with the same procedure. As a whole, the band structure changed only slightly. We show only the behavior in the energy range $\pm 2$eV in Fig.~\ref{fig:pressure-band-dos}. The peak height of the DOS were almost the same, whereas the Fermi level shifted a little toward the middle of the peak by increasing the pressure. This result is consistent with a previous calculation.~\cite{Bianconi-NSM} We confirmed that the DOS peak commonly originates from the saddle loop region $\Omega_{\rm SL}$ defined in Fig.~\ref{fig:H3S-special-path} (see Appendix~\ref{app:D-SL-pressure} for specific data). The Wannier model parameters for different pressures are summarized in Table ~\ref{tab:tbparam}. We observe that the hopping parameters monotonically increases in absolute value by pressure, which is simply due to the compression of the system. Importantly, the farther neighbor hopping parameters $t^{(4)}_{{\rm S}s-{\rm S}s}$ and $t^{(4)}_{{\rm H}s-({\rm H})-{\rm H}s}$ remained positive and sizable regardless of the pressures. One is reminded that the modification of the saddle point into the loop generally requires that the perturbation is stronger than any threshold value (See Fig.~\ref{fig:sc-band}). We confirmed that the single sublattice nearest neighbor hopping model yields only the saddle points and they are modified into the saddle loops by those farther hopping parameters. These results support that the validity of the above statements (i) and (ii) across the experimental pressure range.

The above calculations were performed with the $Im\bar{3}m$ cubic structure, but in the experiments, the crystal structure is thought to suffer from distortions especially at relatively low pressures. For more relations to the experiments, we also calculated the band structure with two types of distortions at 120 GPa. One is the molecular ($R3m$) distortion;~\cite{Duan2014} the three hydrogens are displaced from the center of the bonds so that the local H$_{3}$S molecules are formed [Fig.~\ref{fig:distort-band-dos}(a)]. This structure is more stable than the high symmetry cubic structure in terms of the Born-Oppenheimer energy surface, whereas the high-symmetry structure may be more stable if the quantum nature of the hydrogen positions are considered.~\cite{Errea-Nature2016} The other is the lattice distortion; In a previous experiment,~\cite{Goncharov-PRB2017} a uniform stretch of the crystal was observed in the cubic $(1 1 1)$ direction [Fig.~\ref{fig:distort-band-dos}(b)], though its origin is yet unclear.~\cite{Bianco-PRB2018} To obtain the input crystal structures with these distortions, we performed the variable cell structure optimization for the molecular $R3m$ phase and simply changed the angles between the lattice vectors with the hydrogen positions kept in the middle of the bonds, respectively. In the former case, the lattice parameter was 5.9233 and the cosine of the angle between the lattice vectors, which is exactly zero for the ideal cubic structure, was $\simeq$0.001. The hydrogen positions shifted from the bond centers by about 0.21 bohr in the bonding directions. In the latter, we set the angle cosine to $\simeq$ 0.030 so that the degree of the distortion defined by Ref.~\onlinecite{Goncharov-PRB2017} is 3\%.

We show the first principles band structures and DOS in Fig.~\ref{fig:distort-band-dos} (c) and (d). Although the whole band structure is again consistent, appreciable changes in the band structure and DOS were seen in the vicinity of the Fermi level. The molecular distortion has large effect in particular around the $\Gamma$ point, with which some hole Fermi surfaces disappears. This change is largely responsible for the apparent reduction of the total DOS peak height from the cubic case seen in panel (d), which is also implied by the partial DOS from the region other than the saddle loop one in panel (f). On the other hand, the effect of the lattice distortion is significant on the saddle loop, as indicated at the local band maxima in the $X-M$ and $M-\Gamma$ paths. It lifts the degeneracy at the former maximum and the resulting DOS peak splits into two [panel (d)]. The partial DOS analysis for the saddle loop and other regions [(e) and (f)] shows that the contribution from outside the saddle loop region $D(E)-D_{\rm SL}(E)$ is almost invariant, indicating that the impact of this distortion is local in the ${\bf k}$ space. The origin of this is thought to be the intersublattice coupling [Eq.~(\ref{eq:sub-decouple})] switched on by the reduced symmetry. The DOS peak separation could be observable as a ``pseudogap" structure through any probes that detects the electronic transition processes between the split bands, though its actual magnitude is expected to depend on the degree of the distortion. We note that, apart from the subtle changes of the DOS peak structure, the concentrating trend of the DOS near the Fermi level around the saddle loops is persistent against those distortions. The positive sizable farther neighbor hopping parameters were also obtained as summarized in Table~\ref{tab:tbparam} as well, suggesting the robustness of the mechanism extracted by the sublattice model. These results again validate the statements (i) and (ii) above, as well as clarify the importance of the saddle loops in the experiments.

\section{Conclusions}
In this paper, we have characterized an archetype of the mechanisms that yield the DOS peak in three dimensional crystals. Modifying the second order saddle point (maximum) into local minimum by ``pushing" the band hypersurface, the saddle loop (extremum shell) structure certainly appears and it induces concentration of the DOS at a tiny energy range. Being the high dimensional structure, the saddle loop and extremum shell are difficult to recognize from the standard scheme of visualizing the band structures along linear paths in the Brillouin zone. The existence of the critical points {\it anywhere} (not necessarily at the special points) on these structures are assured by their closed nature, which appear in the DOS spectra as adjacent shoulders. The width of the DOS peak is determined by the band dispersion over the entire loop and shell. We have pointed out that the higher order Lifshitz transition through the energy levels across the DOS peak is a useful indicator to locate those structures in the ${\bf k}$ space. Our theory gives us a deep insight into an important feature, the DOS peaks, of the electronic structure, as well as provides a simple guiding principle for design of electronic materials.

We have demonstrated how the electronic structure in the superconducting H$_{3}$S under pressure is understood with the present theory. The DOS peak in this system, which is thought to be the source of the high temperature superconductivity, is accompanied by several puzzling features such as critical points at apparently fractional low symmetry points. Through the close analysis, we have extracted the saddle loop structure and successfully derived a minimal model for the DOS peak formation. The cubic ReO$_{3}$ structure with sizable positive farther neighbor hopping results in the saddle loops around the $M$ and symmetrically equivalent points, which remains relevant against the lattice doubling. Although we do not go so far as to state that this is the minimal appropriate modeling for the superconductivity of this system, it is confirmed that one of its ingredients---formation of the DOS peak---is more than accidental and the same mechanism can be applicable to various isomorphous systems.

\begin{acknowledgments}
The author thanks to Peter Maksym for fruitful discussions. This work was supported by MEXT Element Strategy Initiative to Form Core Research Center in Japan. Some of the calculations were performed at the Supercomputer Center at the Institute for Solid State Physics in the University of Tokyo. {\sc Vesta}~\cite{Vesta} and {\sc Fermisurfer}~\cite{Fermisurfer} were used for visualization of the data.%was supported as Exploratory Challenge on Post-K computer (Frontiers of Basic Science: Challenging the Limits), and JSPS KAKENHI Grant No. 15K20940 from Japan Society for the Promotion of Science (JSPS). This research used computational resources of the Supercomputer Center at the Institute for Solid State Physics in the University of Tokyo. %This research used computational resources of the K computer provided by the RIKEN Advanced Institute for Computational Science through the HPCI System Research project (Project ID:hp160257, hp170244).
\end{acknowledgments}

\appendix

\section{Spectral functions}
\label{app:spectr}
We calculated the orbital resolved spectral function 
\begin{eqnarray}
A_{i}({\bf k}, \omega)
=
\frac{1}{\pi}{\rm Im}\left[
\frac{1}{\omega - \mathcal{H}({\bf k})-i0^{+}}
\right]_{ii}
,
\end{eqnarray}
with $\mathcal{H}({\bf k})$, $i$ and $0^{+}$ being the Fourier component of the Wannier Hamiltonian, orbital index and infinitesimal positive number. Here, we append the spectral function for the orbital components not displayed in the main text. The sulfur $p_{z}$ and hydrogen H$_{x}$ $s$ and H$_{y}$ $s$ states are irrelevant for the target band responsible for the DOS peak and their components are summed in the figure. All the components show similar behavior among the models, especially for the target band.

\begin{figure*}[t]
    \begin{center}
        \includegraphics[scale=0.60]{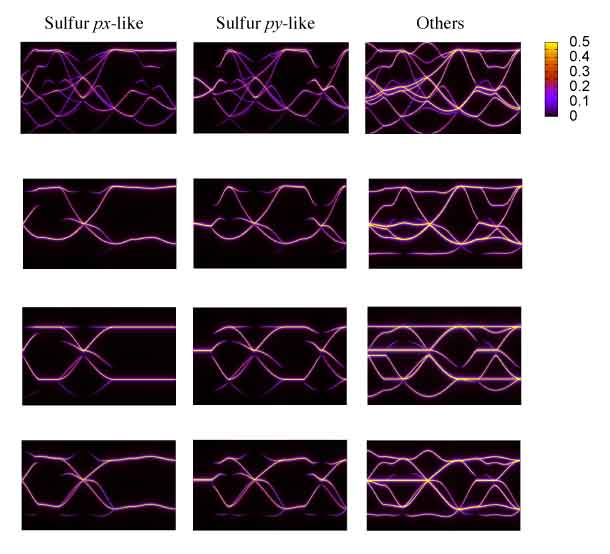}
        \caption{(a)--(d) Spectral functions, continued from Fig.~\ref{fig:Spectr}}
        \label{fig:Spectr-app}
    \end{center}
\end{figure*}

\begin{figure}[h!]
    \begin{center}
        \includegraphics[scale=0.50]{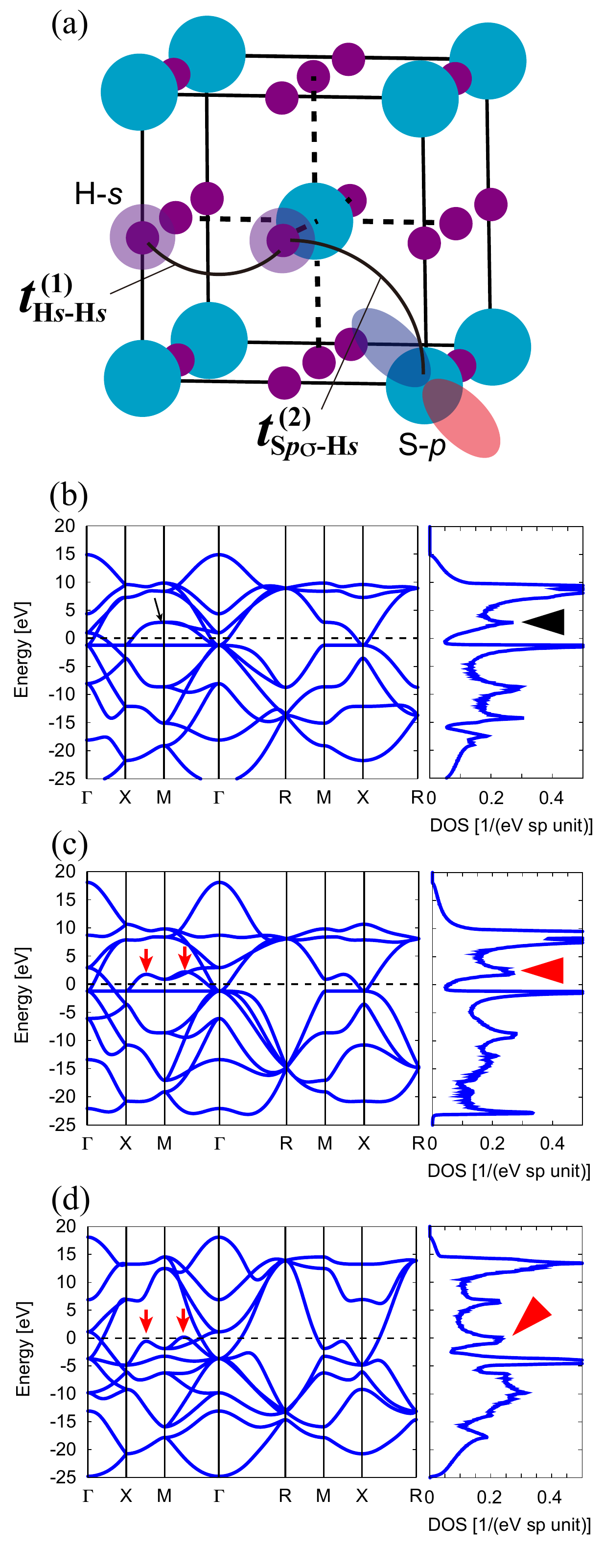}
        \caption{(a) Schematic picture of the intersublattice coupling parameters $t^{(1)}_{{\rm H}s-{\rm H}s}$ and $t^{(2)}_{{\rm S}p\sigma-{\rm H}s}$ in Table~\ref{tab:tbparam-app}. The bold and dashed lines indicate the respective sublattices. (b) The band structure calculated from the nearest neighbor sublattice model [Fig.~\ref{fig:Spectr}(e)] with $t^{(1)}_{{\rm H}s-{\rm H}s}$ and $t^{(2)}_{{\rm S}p\sigma-{\rm H}s}$, (c) that from the ``minimal" model [Fig.~\ref{fig:Spectr}(g)] with $t^{(1)}_{{\rm H}s-{\rm H}s}$ and $t^{(2)}_{{\rm S}p\sigma-{\rm H}s}$ and (d) that from the sublattice model [Fig.~\ref{fig:Spectr}(c)] with $t^{(1)}_{{\rm H}s-{\rm H}s}$ and $t^{(2)}_{{\rm S}p\sigma-{\rm H}s}$.}
        \label{fig:bands-intersub}
    \end{center}
\end{figure}

\section{Effect of intersublattice hopping}
\label{app:intersub}
Among the intersublattice hopping parameters in Table~\ref{tab:tbparam-app}, we found that $t^{(1)}_{{\rm H}s-{\rm H}s}$ and $t^{(2)}_{{\rm S}p\sigma-{\rm H}s}$ [Fig.~\ref{fig:bands-intersub}(a)] have appreciable effect on the band structure, especially when we take them into account concurrently with the farther neighbor hopping parameters $t^{(4)}_{{\rm S}s-{\rm S}s}$ and $t^{(4)}_{{\rm H}s-({\rm H})-{\rm H}s}$. We here summarize their effect with Fig.~\ref{fig:bands-intersub}. In Fig.~\ref{fig:Spectr}(e) and (g) we have shown that the nearest neighbor sublattice hopping model only yields the saddle point and it is modified into the loop by the farther neighbor hopping. If we incorporate the above intersublattice hopping parameters into the nearest neighbor sublattice model, the saddle loop is not formed [panel (b)]. Incorporating the farther neighbor hopping, the saddle loop is eventually formed, where the band dispersion along the $X-M-\Gamma$ path is enhanced than that of our minimal model [panel (c); compared with Fig.~\ref{fig:Spectr}(g)]. Finally, the sublattice model with the full inclusion of the intrasublattice hopping [Fig.~\ref{fig:Spectr}(c)] plus those intersublattice hopping parameters gives us the band structure [panel (d)] which remarkably resembles with the first principles result depicted in Fig.~\ref{fig:H3S-str-band}(c). The DOS peak was commonly formed, though its detailed structure--width, maximum and number of the singularities--depends somehow.

\section{Partial DOS analysis for variable pressures}
\label{app:D-SL-pressure}
\begin{figure}[h!]
    \begin{center}
        \includegraphics[scale=0.45]{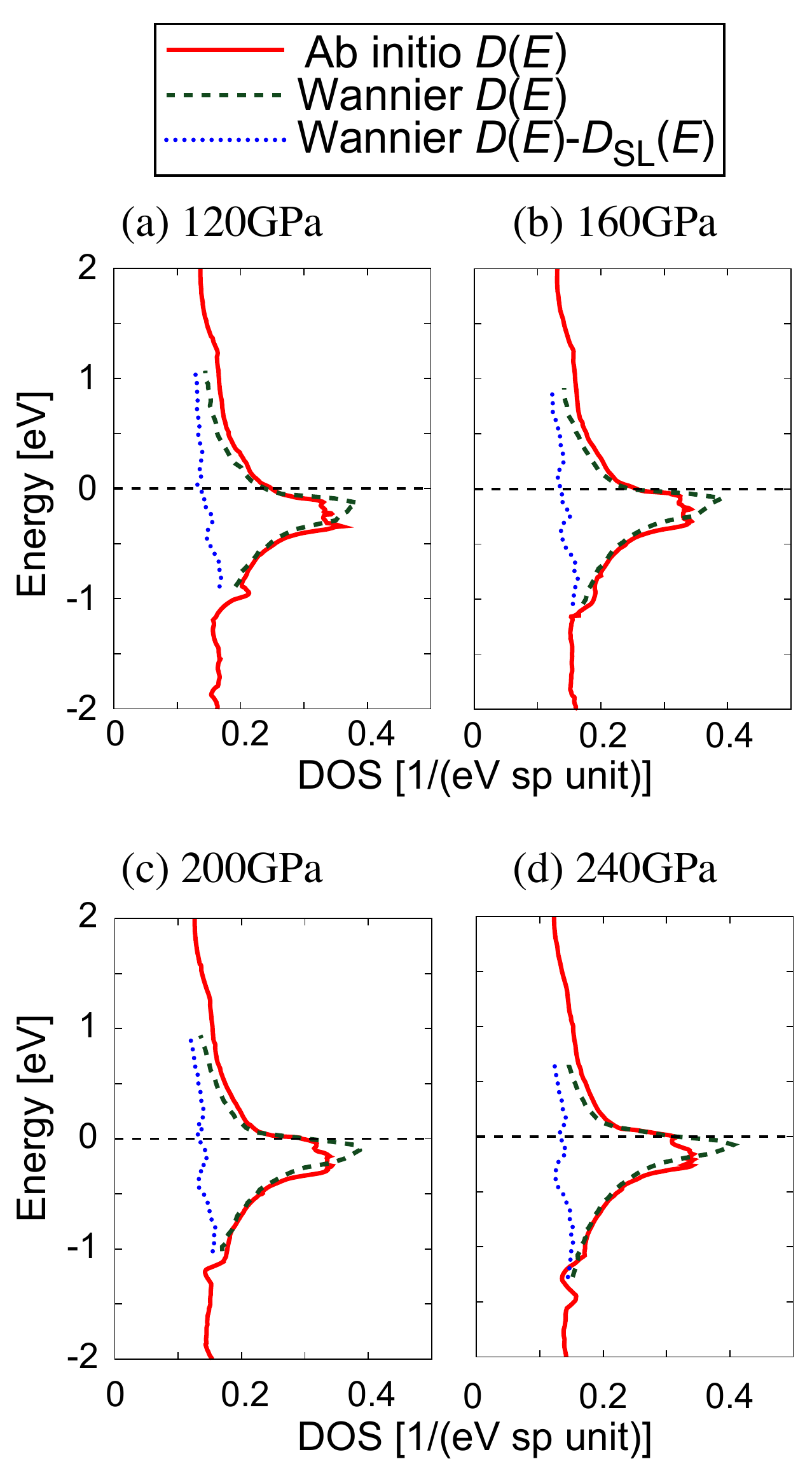}
        \caption{(a)--(d) Total DOS from first principles, from the Wannier models and the partial DOS from the Wannier models, similarly to Fig.~\ref{fig:H3S-special-path} (d).}
        \label{fig:pressure-partialDOS}
    \end{center}
\end{figure}
Here we append the decompositions of the DOS with the partial contribution $D_{\rm SL}(E)$ calculated within the saddle loop region $\Omega_{\rm SL}$ defined in Fig.~\ref{fig:H3S-special-path} for the pressures 120--240GPa. We find that the peaks at the Fermi level are commonly dominated by $D_{\rm SL}(E)$.

\bibliography{reference}

\end{document}